\documentclass[fleqn,usenatbib]{mnras}

\usepackage{newtxtext,newtxmath}
\usepackage[T1]{fontenc}
\usepackage{ae,aecompl}

\usepackage{graphicx}	% Including figure files
\usepackage{amsmath}	% Advanced maths commands
\usepackage[ruled,vlined]{algorithm2e}

\newcommand{\Real}[1]{\mathbb{R}\text{e}\hspace{-1pt}\left[#1\right]}
\newcommand{\Imag}[1]{\mathbb{I}\text{m}\hspace{-1pt}\left[#1\right]}

\title[Redundant-Baseline Calibration of HERA]{Redundant-Baseline Calibration of the Hydrogen Epoch of Reionization Array}

\vspace{-30pt}
\author[J.\ S.\ Dillon et al.]{Joshua~S.~Dillon$^{1,\dagger}$\thanks{Email: jsdillon@berkeley.edu},
Max~Lee$^{1}$,
Zaki~S.~Ali$^{1}$,
Aaron~R.~Parsons$^{1}$,
Naomi~Orosz$^{1}$,
\newauthor
Chuneeta~Devi~Nunhokee$^{1}$,
Paul~La~Plante$^{1,2}$,
Adam~P.~Beardsley$^{3,\dagger}$,
\newauthor
Nicholas~S.~Kern$^{1}$,
Zara~Abdurashidova$^{1}$,
James~E.~Aguirre$^{2}$,
Paul~Alexander$^{4}$,
\newauthor
Yanga~Balfour$^{5}$,
Gianni~Bernardi$^{6,7,5}$,
Tashalee~S.~Billings$^{2}$,
Judd~D.~Bowman$^{3}$,
\newauthor
Richard~F.~Bradley$^{8}$,
Phil~Bull$^{9}$,
Jacob~Burba$^{10}$,
Steve~Carey$^{4}$,
\newauthor
Chris~L.~Carilli$^{11}$,
Carina~Cheng$^{1}$,
David~R.~DeBoer$^{1}$,
Matt~Dexter$^{1}$,
\newauthor
Eloy~de~Lera~Acedo$^{4}$,
John~Ely$^{4}$,
Aaron~Ewall-Wice$^{12}$,
Nicolas~Fagnoni$^{4}$,
\newauthor
Randall~Fritz$^{5}$,
Steven~R.~Furlanetto$^{13}$,
Kingsley~Gale-Sides$^{4}$,
\newauthor
Brian~Glendenning$^{11}$,
Deepthi~Gorthi$^{1}$,
Bradley~Greig$^{14}$,
Jasper~Grobbelaar$^{5}$,
\newauthor
Ziyaad~Halday$^{5}$,
Bryna~J.~Hazelton$^{15,16}$,
Jacqueline~N.~Hewitt$^{12}$,
Jack~Hickish$^{1}$,
\newauthor
Daniel~C.~Jacobs$^{3}$,
Austin~Julius$^{5}$,
Joshua~Kerrigan$^{10}$,
Piyanat~Kittiwisit$^{17}$,
\newauthor
Saul~A.~Kohn$^{2}$,
Matthew~Kolopanis$^{3}$,
Adam~Lanman$^{10}$,
Telalo~Lekalake$^{5}$,
\newauthor
David~Lewis$^{3}$,
Adrian~Liu$^{18}$,
Yin-Zhe~Ma$^{19}$,
David~MacMahon$^{1}$,
Lourence~Malan$^{5}$,
\newauthor
Cresshim~Malgas$^{5}$,
Matthys~Maree$^{5}$,
Zachary~E.~Martinot$^{2}$,
Eunice~Matsetela$^{5}$,
\newauthor
Andrei~Mesinger$^{20}$,
Mathakane~Molewa$^{5}$,
Miguel~F.~Morales$^{15}$,
\newauthor
Tshegofalang~Mosiane$^{5}$,
Steven~Murray$^{3}$,
Abraham~R.~Neben$^{12}$,
Bojan~Nikolic$^{4}$,
\newauthor
Robert~Pascua$^{1}$,
Nipanjana~Patra$^{1}$,
Samantha~Pieterse$^{5}$,
Jonathan~C.~Pober$^{10}$,
\newauthor
Nima~Razavi-Ghods$^{4}$,
Jon~Ringuette$^{15}$,
James~Robnett$^{11}$,
Kathryn~Rosie$^{5}$,
\newauthor
Mario~G.~Santos$^{5,21}$,
Peter~Sims$^{10}$,
Craig~Smith$^{5}$,
Angelo~Syce$^{5}$,
Max~Tegmark$^{12}$,
\newauthor
Nithyanandan~Thyagarajan$^{11,\ddagger}$,
Peter~K.~G.~Williams$^{22,23}$,
Haoxuan~Zheng$^{12}$
\\
$^{1}$ Department of Astronomy, University of California, Berkeley, CA\\
$^{2}$ Department of Physics and Astronomy, University of Pennsylvania, Philadelphia, PA\\
$^{3}$ School of Earth and Space Exploration, Arizona State University, Tempe, AZ\\
$^{4}$ Cavendish Astrophysics, University of Cambridge, Cambridge, UK\\
$^{5}$ South African Radio Astronomy Observatory, Black River Park, 2 Fir Street, Observatory, Cape Town, 7925, South Africa\\
$^{6}$ Department of Physics and Electronics, Rhodes University, PO Box 94, Grahamstown, 6140, South Africa\\
$^{7}$ INAF-Istituto di Radioastronomia, via Gobetti 101, 40129 Bologna, Italy\\
$^{8}$ National Radio Astronomy Observatory, Charlottesville, VA\\
$^{9}$ Queen Mary University London\\
$^{10}$ Department of Physics, Brown University, Providence, RI\\
$^{11}$ National Radio Astronomy Observatory, Socorro, NM\\
$^{12}$ Department of Physics, Massachusetts Institute of Technology, Cambridge, MA\\
$^{13}$ Department of Physics and Astronomy, University of California, Los Angeles, CA\\
$^{14}$ School of Physics, University of Melbourne, Parkville, VIC 3010, Australia\\
$^{15}$ Department of Physics, University of Washington, Seattle, WA\\
$^{16}$ eScience Institute, University of Washington, Seattle, WA\\
$^{17}$ School of Chemistry and Physics, University of KwaZulu-Natal, Westville Campus, Private Bag X54001, Durban, South Africa\\
$^{18}$ Department of Physics and McGill Space Institute, McGill University, 3600 University Street, Montreal, QC H3A 2T8, Canada\\
$^{19}$ School of Chemistry and Physics, University of KwaZulu-Natal, Westville Campus, Durban, 4000, South Africa\\
$^{20}$ Scuola Normale Superiore, 56126 Pisa, PI, Italy\\
$^{21}$ Department of Physics and Astronomy,  University of Western Cape, Cape Town, 7535, South Africa\\
$^{22}$ Center for Astrophysics, Harvard \& Smithsonian, Cambridge, MA\\
$^{23}$ American Astronomical Society, Washington, DC\\
$^{\dagger}$ NSF Astronomy and Astrophysics Postdoctoral Fellow \hspace{5pt} $^{\ddagger}$ NRAO Jansky Fellow
}

\date{\today}%Accepted XXX. Received YYY; in original form ZZZ}

\pubyear{2019}

\begin{document}
\label{firstpage}
\pagerange{\pageref{firstpage}--\pageref{lastpage}}
\maketitle

\begin{abstract}
In 21\,cm cosmology, precision calibration is key to the separation of the neutral hydrogen signal from  very bright but spectrally-smooth astrophysical foregrounds. The Hydrogen Epoch of Reionization Array (HERA), an interferometer specialized for 21\,cm cosmology and now under construction in South Africa, was designed to be largely calibrated using the self-consistency of repeated measurements of the same interferometric modes. This technique, known as \emph{redundant-baseline calibration} resolves most of the internal degrees of freedom in the calibration problem. It assumes, however, on antenna elements with identical primary beams placed precisely on a redundant grid. In this work, we review the detailed implementation of the algorithms enabling redundant-baseline calibration and report results with HERA data. We quantify the effects of real-world non-redundancy and how they compare to the idealized scenario in which redundant measurements differ only in their noise realizations. Finally, we study how non-redundancy can produce spurious temporal structure in our calibration solutions---both in data and in simulations---and present strategies for mitigating that structure. 
\end{abstract}


%%%%%%%%%%%%%%%%%%%%%%%%%%%%%%%%%%%%%%%%%%%%%%%%%%

%%%%%%%%%%%%%%%%% BODY OF PAPER %%%%%%%%%%%%%%%%%%

\section{Introduction}

21\,cm cosmology, the tomographic mapping of the redshifted hyperfine transition of neutral hydrogen, has the potential to provide direct access to the majority of the luminous matter in the universe for the first time \citep{furlanetto_et_al2006, morales_wyithe2010, pritchard_loeb2012, Zaroubi2013, Loeb_and_Furlanetto2013, Liu_Shaw2019, Furlanetto_et_al2019c, Liu_et_al2019, Burns_et_al2019}. At lower redshifts, 21\,cm tomography can enable intensity mapping of self-shielded gas within galaxies, allowing for precise measurements of baryon acoustic oscillations across cosmic time \citep{Chang_et_al2008, Loeb_and_Wyithe2008, Cosmic_Visions2018, Kovetz_et_al2019, Slosar_et_al2019}. At higher redshifts, 21\,cm cosmology promises a new window on the ``Cosmic Dawn'', spanning from the first stars through to the Epoch of Reionization (EoR) by probing the intergalactic medium's (IGM) temperature, density, and ionization state \citep{Furlanetto_et_al2019b, Mirocha_et_al2019, Chang_et_al2019, Furlanetto_et_al2019a, Alvarez_et_al2019}. Both the sky-averaged 21\,cm brightness temperature and its fluctuations encode key information about these processes. With the as-yet-unconfirmed detection of a surprisingly strong global absorption signal at $z \approx 17$ by the EDGES team \citep{bowman_et_al2018}, there is pressing need for followup observations of both the global 21\,cm signal and its fluctuations during the Cosmic Dawn.

The primary challenge of 21\,cm cosmology across redshifts is distinguishing 21\,cm signal from comparatively nearby astrophysical foregrounds, namely the continuum emission from our Galaxy and other radio-bright galaxies, that are $\sim$10$^5$ times brighter. In 21\,cm tomography with radio interferometers---the focus of this work---that separation relies on the spectral smoothness of foregrounds compared to the complex spectral structure of the 21\,cm signal where different frequencies correspond to distinct regions of the IGM. That separation is complicated by the inherently chromatic nature of interferometric measurements. With an ideal instrument that contamination is limited in to a wedge-shaped region of Fourier space \citep{Datta_2010,vedantham_2012,parsons_et_al2012a,parsons_et_al2012b, liu_et_al2014a, liu_et_al2014b}, leaving the remaining modes clean for a detection and characterization of the 21\,cm signal via its power spectrum. 

However, any unmodeled effects that result in additional spectral structure in the instrument response risk destroying that clean separation in Fourier space. Chief among these is the bandpass function of each antenna's signal chain, which multiplies the true per-antenna voltages in our measured visibilities. This effect can be modeled as a complex per-antenna and per-polarization gain as a function of frequency and time, namely
\begin{equation}
    V_{ij}^\text{obs}(\nu,t) = g_i(\nu,t) g_j^*(\nu,t) V_{ij}^\text{true}(\nu,t) + n_{ij}(\nu,t), \label{eq:gains}
\end{equation}
where $V_{ij}$ is the visibility measured for the baseline between antennas $i$ and $j$, $g_i$ is the gain on the $i$th antenna, and $n_{ij}$ is the Gaussian-distributed thermal noise in that measurement. The process of correcting for these gains is often called \emph{direction-independent calibration} to distinguish it from the problem of accounting for the spatial response of each antenna element. In this paper, we will just use \emph{calibration} as a shorthand.\footnote{We also ignore the so-called $D$-terms, which mix polarization responses, which are likely much smaller than the relevant calibration errors and have less impact on the calibration of visibilities between antennas of the same polarization---the ones most sensitive to unpolarized 21\,cm cosmological signal.}  
Errors in calibration that produce spectral structure in the effective instrument response, when multiplied by the overwhelmingly bright foregrounds, can completely contaminate otherwise clean Fourier modes. For example, \citet{ewall-wice_et_al2016-EoXLimits} saw that sub-percent-level cable reflections produce sinusoidal ripples in the bandpass that create attenuated but still very bright copies of the foreground wedge centered at the line-of-sight cosmological mode, $k_\|$, corresponding to the reflection's delay.

Traditionally, the calibration of radio interferometers requires precise models of the sky and antenna beams to simulate the expected $V_{ij}$ for each antenna pair and thus solve for each $g_i$ as part of a large, over-constrained system of equations. Sky-based calibration is especially difficult for arrays optimized for 21\,cm cosmology, which often feature large fields of view and limited steerability of elements. Worse, models of continuum foregrounds---especially, diffuse and polarized emissions---are rarely accurate to better than the percent-level. Unmodeled foregrounds, even those below the confusion limit of current and upcoming arrays, can produce ruinous spectral calibration errors \citep{barry_et_al2016}. This effect can be mitigated by calibrating with only the shortest, least spectrally-complex baselines \citep{ewall-wice_et_al2017}, though this may make the calibration less accurate as it relies more heavily on the poorly-modeled diffuse Galactic emission. Alternatively, one can impose \emph{a priori} constraints on calibration solutions, either by the consensus optimization technique \citep{Yatawatta2015, Yatawatta2016} used in \citet{patil_et_al2017} and \citet{Mertens_et_al2020}, via low-order polynomials \citep{Barry_et_al2019, Barry_et_al2019b}, or by directly filtering gains \citep{kern_et_al2019c}. This general approach relies on the inherent spectral smoothness smoothness of the instrument response, including its complex, per-antenna gains. Much work has been done with simulations \citep{ewallwice_et_al2016, Trott_et_al2017} and field measurements \citep{patra_et_al2018} to verify the smoothness and calibratability of arrays. It remains to be demonstrated in which regimes the \emph{a priori} assumption of spectral smoothness will hold for real-world antennas operating inside large arrays with complex electromagnetic interactions.

An alternate approach to calibrating arrays with many pairs of antennas with the same physical separation is to solve for the gains and unique visibilities simultaneously, using no prior information about the sky or the instrument other than the assumption that elements are identical and placed correctly. This approach, called \emph{redundant-baseline calibration}, works well when the number of unique baseline separations is much smaller than the total number of measurements, as is usually the case when the array is constructed on a regular grid. This approach, developed in \citet{wieringa1992} and formalized in \citet{liu_et_al2010}, simplifies the problem by solving for most of calibration's degrees of freedom---one complex antenna gain per antenna and one complex visibility per unique baseline type for each frequency and polarization---using only the internal consistency of redundant measurements. 

However, redundant calibration cannot resolve a small number of degenerate parameters---four per frequency and per polarization---and must ultimately reference the sky to resolve them \citep{zheng_et_al2014, dillon_et_al2017}, a process we call \emph{absolute calibration}, since its primary purpose is to set a flux scale and pointing center. Just as modeling errors plague sky-based calibration, redundant-baseline calibration can suffer from similar problems due to position errors and beam-to-beam variation between antenna elements \citep{Orosz_et_al2019} or to sky modeling errors in the absolute calibration step \citep{Byrne_et_al2019}. Likewise, these too can be mitigated by \emph{a priori} constraints on bandpass spectral smoothness or by using only short baselines \citep{Orosz_et_al2019}.  Hybrid techniques that transition from redundant-baseline to sky-based calibration with increasing sky and instrument knowledge are also being explored \citep{Sievers2017}.

Different first-generation arrays aiming to detect the 21\,cm power spectrum were constructed to take advantage of different techniques. The LOw-Frequency ARray (LOFAR; \citealt{van_haarlem_et_al2013, Mertens_et_al2020}) and Phase I of the Murchison Widefield Array (MWA; \citealt{tingay_et_al2013,bowman_et_al2012, Trott_et_al2020}) were both designed with minimal baseline redundancy to optimize for a $uv$-coverage and thus the ability to image the sky and iteratively calibrate off that image. By contrast, the technology demonstrator MITEoR \citep{zheng_et_al2014, Zheng_et_al2017} and the Donald C.\ Backer Precision Array for Probing the Epoch of Reionization (PAPER; \citealt{parsons_et_al2010,ali_et_al2015}) were built on regular grids which both enable redundant-baseline calibration and focus sensitivity on the measurement of a few modes for a delay-spectrum analysis \citep{parsons_et_al2012b}. Phase II of the MWA added some elements on a regular grid, enabling a comparison of both types of calibration \citep{Li_et_al2018}, as well as hybrid approaches.

As a second generation instrument, the Hydrogen Epoch of Reionization Array (HERA; \citealt{deboer_et_al2017}) borrows approaches from its predecessors. When it is complete, HERA will consist of 350 14\,m parabolic dishes observing at 50--250\,MHz ($4.7 \lesssim z \lesssim 27$). Of these, 320 are packed hexagonally into a dense core while the remaining 30 outrigger antennas provide longer baselines ($\sim$1\,km) for improved imaging. HERA was designed such that the entire array could be redundantly calibrated, including the outriggers \citep{dillon_parsons2016}. It also features a core split into three offset sectors to create denser simultaneous $uv$-coverage that improves HERA's ability to map diffuse galactic structure \citep{Dillon_et_al2015a}. 

In this work, we describe results from the redundant-baseline of Phase I of HERA. In Phase I, HERA existed as a hybrid of PAPER and the final HERA system; it featured modified PAPER dipole feeds suspended over HERA dishes \citep{neben_et_al2016, ewallwice_et_al2016, patra_et_al2018, Fagnoni_et_al2019a} and used the PAPER signal chain and correlator. The legacy PAPER components are all now being replaced as construction on the full HERA system continues through 2020. This work complements other recent work with HERA on both systematics mitigation \citep{kern_et_al2019a, kern_et_al2019b} and sky-based calibration and (post-redundant) absolute calibration \citep{kern_et_al2019c}. Likewise, this paper complements the investigation of the spectral smoothness of HERA's calibration solutions in \citet{kern_et_al2019c} with an investigation of the temporal structure of those solutions and its origin. Along with those papers, it is meant to lay the groundwork for forthcoming HERA Phase I upper-limits on the 21\,cm power spectrum.

HERA is also an important testbed for redundant-baseline calibration. Next-generation arrays across redshifts are being considered which rely on Fast Fourier Transform (FFT) correlation \citep{tegmark_zaldarriaga2009, tegmark_zaldarriaga2010} to make arrays with large numbers of elements feasible \citep{Cosmic_Visions2018, Slosar_et_al2019, Ahmed_et_al2019, HERA_2020_White_Paper}. FFT-correlation, which can reduce the cost-scaling of correlating $N$-antenna arrays from $\mathcal{O}(N^2)$ to $\mathcal{O}(N\log N)$, achieves that speed-up via a form of data compression that relies on precise relative calibration of antennas---precisely the terms solved for by redundant-baseline calibration---in real time. While not the only route to faster correlation \citep{Morales2011, Thyagarajan_et_al2017, Kent_et_al2019}, redundant-baseline calibration is likely a necessary enabling technology for futuristic 21\,cm interferometers.

Previous work with PAPER \citep{ali_et_al2015, kolopanis_et_al2019} and MITEoR \citep{zheng_et_al2014, Zheng_et_al2017} showed both the promise of redundant-baseline calibration and its ability to clearly spot deviations from ideal behavior. However, HERA's high sensitivity is an opportunity to assess redundant-baseline calibration in a more systematic way. How redundant is HERA and how do we quantify redundancy? In this paper, we explore several ways of answering that question using 18 days of observation with HERA Phase I (Section~\ref{sec:hera_redundancy}). Along the way we review redundant-baseline calibration, elaborating on previously unpublished implementation details and a corrected and expanded exploration of one of the key metrics of redundancy, $\chi^2$ (Section~\ref{sec:redcal}). Finally, we explore how the observed temporal structure in our calibration solutions can be understood as a consequence of non-redundancy (Section~\ref{sec:temporal}), complementing the exploration of spectral structure in calibration solutions in \citet{kern_et_al2019c}.

%%%%%%%%%%%%%%%%%%%%%%%%%%%%%%%%%%%%%%%%%%%%%%%%%%%%%%%%%%%%%%%%%%%%%%%%%%%%%%%%%%%%%%%%%%

\section{The Theory and Practice of Redundant-Baseline Calibration} \label{sec:redcal}

In this section, we review the mathematical underpinnings (Section~\ref{sec:redcal_review}) and practical algorithmic implementation (Section~\ref{sec:redcal_implementation}) of redundant-baseline calibration. While some of this material has been previously published \citep{liu_et_al2010, zheng_et_al2014, Zheng_et_al2017, dillon_et_al2017, Li_et_al2018}, a number of the implementation details are missing from the literature, especially the \texttt{firstcal} (Section~\ref{sec:firstcal}) and  \texttt{omnical} algorithms (Section~\ref{sec:omnical}), and the proper counting of degrees of freedom for normalizing $\chi^2$ (Section~\ref{sec:chisq_dof}).

In implementing and refining the technique for HERA, we have strived to maintain the independence of our techniques from any detailed knowledge about the sky or the array. All we need need to know about the array is where the antenna elements are and that they are approximately identical to one another. From that point, we can perform \emph{most} of the calibration and learn quite a bit about how well the array is functioning. This has proven especially useful for HERA, since for both Phase I and Phase II we have commissioned effectively brand-new arrays.

\subsection{Redundant-Baseline Calibration Review} \label{sec:redcal_review}

Fundamentally, redundant-baseline calibration is a process for finding a solution to a system of equations of the form
\begin{equation}
    V_{ij}^\text{obs} = g_i g_j^* V_{i-j} \label{eq:redcal_equation}
\end{equation}
that minimizes $\chi^2$ defined as
\begin{equation}
    \chi^2 \equiv \sum_{i < j} \frac{\left| V_{ij}^\text{obs} - g_i g_j^* V_{i-j} \right|^2}{\sigma_{ij}^2}. \label{eq:chisq}
\end{equation}
Here $V_{i-j}$ is a shorthand for our estimate of the true visibility with the same baseline separation as the one between antennas $i$ and $j$, using the fact that $V_{ij}^\text{true}$ depends only on that separation vector. Ideally, $\langle V_{i-j} \rangle = V_{ij}^\text{true}$, but in the real world both non-redundancy and degeneracies make them differ from one another. $\sigma_{ij}^2$ is the variance of $n_{ij}$ in Equation~\ref{eq:gains}. For simplicity, we have dropped the explicit dependence on time and frequency, though all these terms are, in principle, functions of both. Solving the system of equation generally requires linearizing Equation~\ref{eq:redcal_equation}. Originally, \citet{wieringa1992} proposed taking the logarithm of both sides and then solving for the real and imaginary parts separately as two linear systems of equations. \citet{liu_et_al2010} showed that this \texttt{logcal} procedure yields biased results and instead proposed a \texttt{lincal} approach using the iterative Gauss-Newton algorithm, Taylor expanding around approximate solutions and updating the solutions, to first order, using a linear system of equations. Similar methods have been employed in the literature, with different non-linear optimization algorithms such as Levenberg–Marquardt \citep{Grobler_et_al2018}. For a pedagogical review, see \citet{dillon_et_al2017}. Later in this section, we will detail an alternative iterative algorithm that avoids matrix inversion. This \texttt{omnical} algorithm was originally developed for and used in \citet{zheng_et_al2014}, but was never explained in the literature.

After minimizing $\chi^2$ by whatever method, the degrees of freedom in calibration that leave $\chi^2$ unchanged are the \emph{degeneracies} of the system of equations. When polarizations are calibrated independently (as is the case in this work), there are four such degeneracies per polarization and frequency. These are the overall amplitude ($g_i \rightarrow Ag_i$,  $V_{i-j} \rightarrow A^{-2} V_{i-j}$), the overall phase ($g_i \rightarrow e^{i\psi}g_i$), the East-West tip-tilt ($g_i \rightarrow g_i e^{i\Phi_x x_i}$, $V_{i-j} \rightarrow V_{i-j} e^{-i\Phi_x \Delta x_{ij}}$), and the North-South tip-tilt ($g_i \rightarrow g_i e^{i\Phi_y y_i}$, $V_{i-j} \rightarrow V_{i-j} e^{-i\Phi_y \Delta y_{ij}}$) where $A$, $\psi$, $\Phi_x$, and $\Phi_y$ are arbitrary real scalars and $x_i$ and $y_i$ are antenna position components. While three these can be solved by absolute calibration using a sky model as a reference, the overall phase cannot because it is merely an arbitrary convention with no physical significance.

\subsection{Practical Implementation of Redundant-Baseline Calibration} \label{sec:redcal_implementation}

In practice, redundant-baseline calibration must be performed as a series of iterative steps, each bringing us closer to a solution that minimizes $\chi^2$. Without a good starting point, phase wrapping issues plague \texttt{logcal} while \texttt{lincal} and \texttt{omnical} converge slowly, if at all \citep{zheng_et_al2014, Joseph_et_al2018}. Getting ``good enough'' gain phases is key to achieving convergence and avoiding the introduction of spectral structure in the degeneracies \citep{dillon_et_al2017}.  In this section, we explain our refined method for implementing these steps, starting with \texttt{firstcal}, a sky-independent way to find a starting point for the rest of redundant-baseline calibration.

\subsubsection{Overall phase and delay calibration: \texttt{firstcal}} \label{sec:firstcal}

At a given time, any gain can be written without loss of generality as 
\begin{equation}
    g_j(\nu) = A_j(\nu) e^{i\phi_j(\nu) + 2\pi i \nu \tau_j  + i\theta_j}
\end{equation}
where $A$, $\phi$, $\tau$,and $\theta$ are all real. Most of the spectral structure in a gain's phase comes the delay, $\tau_j$, corresponding to the light travel time from the antenna to the correlator. Differences in delays are often driven by small differences in cable length. While an overall delay added to all antennas has no effect on the measured visibilities, delay differences are a key factor to correct for. We also include an overall phase term, $\theta_j$, to account for antenna feeds accidentally installed with a 180$^\circ$ rotation. Both could be absorbed into a general frequency-dependent phase, $\phi_j(\nu)$, but it is useful for what follows to separate them out. In past work this initial phase calibration is accomplished by a ``rough calibration'' referenced to the sky \citep{zheng_et_al2014, ali_et_al2015}. However, we have pursued an alternate approach, which can be performed completely independently (and in parallel) for each integration without reference to a sky-model.\footnote{In practice, we do not expect $\tau$ and $\theta$ to vary significantly over the course of a night, but since this step is not rate-limiting computationally, we find it useful to perform repeatedly both to make calibration more trivially parallelizable and to provide additional data quality checks.}

Our approach, \texttt{firstcal}, uses redundancy to solve for one delay and phase per antenna and per polarization (but not per frequency), up to a set of degeneracies that turn out to be a subset of those listed above.\footnote{While originally developed for PAPER and HERA, a simpler, delay-only variant of \texttt{firstcal} first appeared in the literature in \citet{Li_et_al2018} and was applied to MWA data.} The key idea is to look at pairs of measured visibilities, $V_{ij}$ and $V_{kl}$, that probe the same redundant baseline, $V_{i-j}$:
\begin{align} \label{eq:firstcal}
    \frac{V_{ij}V^*_{kl}}{\left|V_{ij}\right| \left|V_{kl} \right|} 
    & \approx \frac{g_i g_j^* V_{i-j} g_k^* g_l V^*_{i-j}}{\left|g_i\right|\left|g_j\right|\left|g_k\right|\left|g_l\right|\left|V_{i-j}\right|^2} \nonumber \\ 
    & \approx \exp \left[i\left(\theta_i-\theta_j-\theta_k+\theta_l\right) + 2\pi i \nu \left(\tau_i - \tau_j - \tau_k + \tau_l \right) \right].
\end{align}
Here we have neglected any frequency-dependence of the gain phases not captured by a delay term. Normalizing by the magnitude of the visibilities (as opposed to taking the ratio of visibilities) has the added benefit of reducing the added weight that would otherwise be assigned to channels with high levels of contamination from radio-frequency interference (RFI).

Each pair of baselines within any given redundant-baseline group gives us another equation in the form of Equation~\ref{eq:firstcal} that involves (at most) four antennas. To linearize this set of equations, we perform delay and phase estimation on the left-hand side using the FFT-based Quinn's Second Estimator (\citealt{Quinn1997}; see Appendix~\ref{sec:fft_dly}). This becomes two large systems of equations (one for the $\theta$ terms and one for the $\tau$ terms), though one could likely use a subset of the redundant-baseline groups to find a satisfactory starting point for full redundant-baseline calibration.

While solving for the delay terms is fairly straightforward, solving for $\theta$ terms is complicated by phase wrapping, since either side of Equation~\ref{eq:firstcal} can have an arbitrary $\pm 2\pi N$.  We find that repeated iterations of the \texttt{firstcal} algorithm converge to a stable solution that reduces $\chi^2$ in Equation~\ref{eq:chisq} considerably.\footnote{Strictly speaking, at this point, we only have estimators for the gains but not the unique baseline visibilities. For the purpose of calculating $\chi^2$, $V_{i-j}$ is estimated by averaging calibrated visibilities within each group. This is suboptimal when different baselines have different noise levels.} 

\begin{figure}
    \centering
    \includegraphics[width=.5\textwidth]{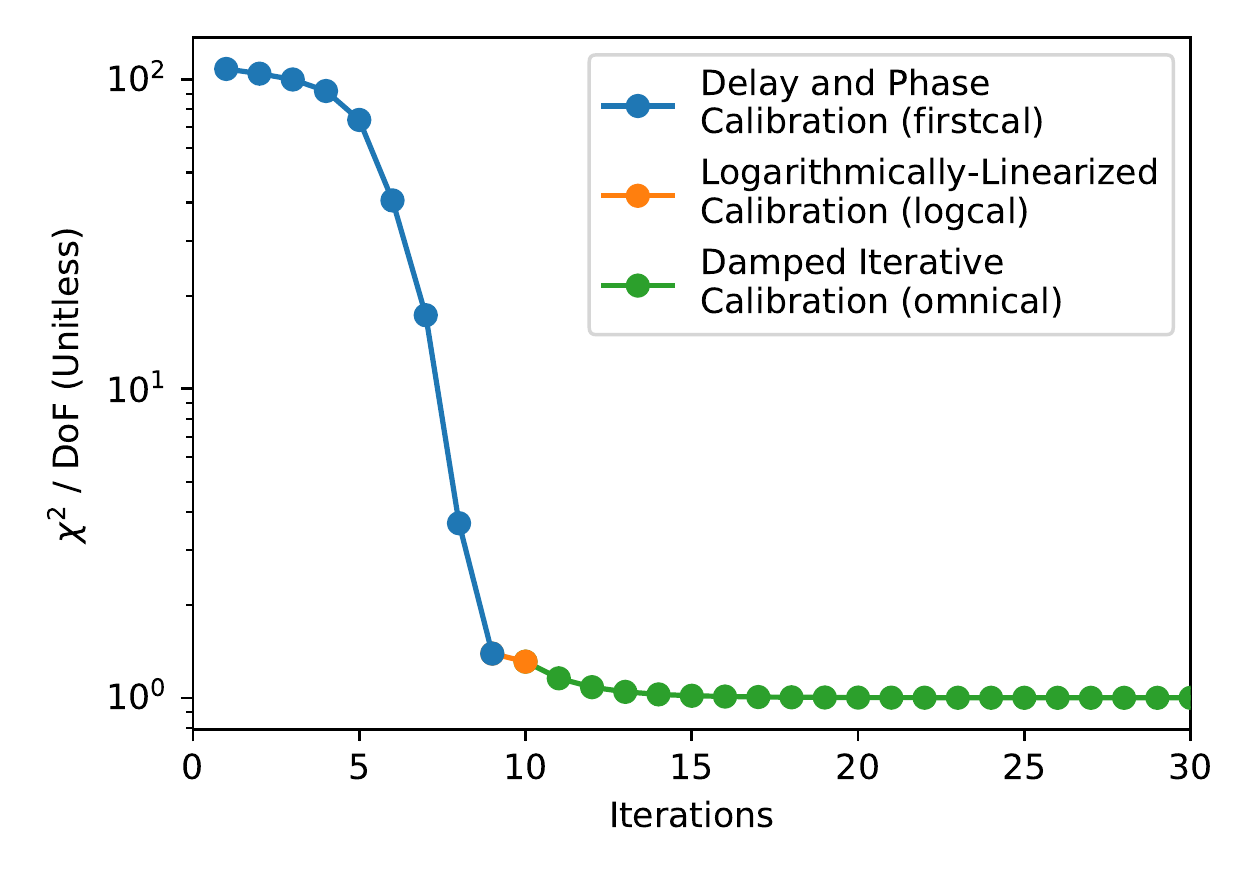}
    \vspace{-10pt}
    \caption{The three stages of redundant calibration---\texttt{firstcal}, \texttt{logcal}, and \texttt{omnical}---converge rapidly to a minimized value of $\chi^2$. Here we show the mean value of $\chi^2$ per degree of freedom (DoF), which has an expectation value of 1, calculated at each step in the calibration of a simulated 37-element array. The simulation details are in Section~\ref{sec:firstcal}. For more on the calculation of DoF, see Section~\ref{sec:chisq_dof}. In \texttt{firstcal}, where $V_{i-j}$ is not directly solved for, an average of calibrated visibilities within a redundant group is used. It should be noted that $x$-axis here is a bit misleading; \texttt{logcal} is not iterative and the computational cost of \texttt{firstcal} and \texttt{omnical} depends on the observation and are not always directly comparable.}
    \label{fig:chisq_convergence}
\end{figure}

To demonstrate this technique, we simulated a redundant 37-element hexagonal array from 100--200\,MHz with relatively realistic, frequency-dependent complex gains, and added random overall phases between $0$ and $2\pi$ and random delays between -20 and 20\,ns. Our visibilities are random, rather than drawn from a sky model, but they are perfectly redundant after calibration. We then add complex Gaussian random noise drawn such that calibrated visibilities have identical noise variance, as would be the case with pure sky-noise \citep{thompson_et_al2017}. We set the signal-to-noise ratio (SNR) on the visibilities to approximately 10, roughly consistent with typical HERA observations.
In Figure~\ref{fig:chisq_convergence}, we show \texttt{firstcal}'s ability to converge in a relatively small number of iterations, yielding a good starting point for subsequent calibration that allows rapid minimization of $\chi^2$ despite the high dimensionality of the problem. 

\subsubsection{Logarithmically linearized redundant-baseline calibration: \texttt{logcal}}

The next step in redundant-baseline calibration, \texttt{logcal}, linearizes Equation~\ref{eq:redcal_equation} by taking the logarithm of both sides. This technique has been extensively reviewed in the literature \citep{wieringa1992, liu_et_al2016, zheng_et_al2014, ali_et_al2015, dillon_et_al2017, Li_et_al2018} and we use it here without further refinement. However, we will briefly review the key formalism here since it will prove useful when we return to DoF counting in Section~\ref{sec:chisq_dof}.

If we define our complex gains as $g_j \equiv \exp\left[\eta_j + i\varphi_j \right]$, where $\eta_j$ and $\varphi_j$ are both real, and if we take the natural logarithm of both sides of Equation~\ref{eq:redcal_equation} and then break it apart into real and imaginary terms, we get
\begin{align}
    \Real{\ln{V^\text{obs}_{ij}}} &= \eta_i + \eta_j + \Real{\ln{V_{i-j}}} \text{ and} \nonumber \\
    \Imag{\ln{V^\text{obs}_{ij}}} &= \varphi_i - \varphi_j + \Imag{\ln{V_{i-j}}}. \label{eq:logcal}
\end{align}
This produces two decoupled systems of linear equations, which we can write as
\begin{align}
    \Real{\mathbf{d}} &= \mathbf{A} \mbox{ }\Real{\mathbf{x}}  \text{ and} \nonumber \\
    \Imag{\mathbf{d}} &= \mathbf{B} \mbox{ }\Imag{\mathbf{x}} \label{eq:logcal_matrices}
\end{align}
where $\mathbf{d}$ is a vector of the natural logarithms of the observed visibilities and $\mathbf{x}$ includes the gains and visibility solutions.  The matrices $\mathbf{A}$ and $\mathbf{B}$ encode the coefficients in Equation~\ref{eq:logcal}; every entry is either 1, 0, or in the case of $\mathbf{B}$, -1. These systems of equations can be solved with the standard linear least-squares estimators, however the solution is biased. While it may be a step in the right direction, it will generally not yield the absolute minimum possible value of $\chi^2$, as we see in Figure~\ref{fig:chisq_convergence} \citep{liu_et_al2010}.

\subsubsection{Damped fixed-point iteration redundant-baseline calibration: \texttt{omnical}} \label{sec:omnical}

\citet{liu_et_al2010} introduced an alternative approach to \texttt{logcal} that produced unbiased results using the Gauss-Newton algorithm. Instead of taking the logarithm, we express gains as $g_i = g_i^0 + \Delta g_i$ and unique visibility solutions as $V_{i-j} = V_{i-j}^0 + \Delta V_{i-j}$. Plugging that into Equation~\ref{eq:redcal_equation} and dropping second order terms, this yields
\begin{equation}
    V_{ij}^\text{obs} - g_i^0 g_j^{0*} V_{i-j}^0 = g_j^{*0}V_{i-j}^0 \Delta g_i + g_i^0 V_{i-j}^0 \Delta g_j^* + g_i^0 g_j^{*0}\Delta V_{i-j}.
\end{equation}
The \texttt{lincal} algorithm simply solves for the $\Delta$-terms with a standard noise-weighted least-squares optimization, updates, and repeats until convergence. 

However, a faster method called \texttt{omnical} was developed for \citet{zheng_et_al2014} and used in \citet{ali_et_al2015} without an explicit acknowledgement that it was, in fact, a different algorithm. The key idea of \texttt{omnical} is to update each gain and visibility as if all other gains and visibilities were constant. This technique is essentially a form of the method for solving non-linear systems of equations via fixed-point iteration. Ideally, as our model of the data improves with each subsequent iteration, the statement that
\begin{equation}
    V^\text{obs}_{ij} \approx g_i^n g_j^{n*} V_{i-j}^n
\end{equation}
becomes closer and closer to accurate (where the superscripts denote the $n$ iteration). It follows then that
\begin{equation}
    g_i \approx V^\text{obs}_{ij} / (g_j^{n*} V_{i-j}^n).
\end{equation}
Since this equation should hold for all $j$, we can use it to update $g_i$ by holding fixed all other variables in the system. Thus, to update $g_i$, we take an average over baselines that include antenna $i$ with weights $w_{ij}$, giving us
\begin{align}
    g_i' &= \left(\sum_j \frac{w_{ij} V^\text{obs}_{ij}}{g_j^{n*} V_{i-j}^n}\middle) \right/ \sum_j w_{ij} \nonumber \\
         &= g_i^n \left(\sum_j w_{ij} V^\text{obs}_{ij} / y_{ij}^n\middle) \right/ \sum_j w_{ij}
\end{align}
where $y_{ij}^n \equiv g_i^n g_j^{n*} V_{i-j}^n$.

This method for iteratively updating gains (and analogously for visibility solutions) is detailed in Algorithm~\ref{alg:omnical}.  
\begin{algorithm}[]
\SetAlgoLined
% \KwResult{Write here the result }
 Generate initial gain and unique visibility solutions, $g_i^0$ and $V_{i-j}^0$ via e.g.\ \texttt{firstcal} and \texttt{logcal}\;
 \For{$0 \leq n < N_{\rm max}$}{
%  \While{}{
  Evaluate all $y_{ij}^n = g_i^n g_j^{n*} V_{i-j}^n$\;
  Update weights $w_{ij} = \left(y_{ij}^n\right)^2 / \sigma_{ij}^2$\;
  \For{$g_i^n \in \mathbf{g}^n$}{
    $g_i' = g_i^n \left(\sum_j w_{ij} V_{ij}^\text{obs} / y_{ij}^n \right) / \sum_j w_{ij}$\;
    $g_i^{n+1} = (1-\delta)g_i^n + \delta g_i'$\;
  }
  \For{$V_{i-j}^n \in \mathbf{V}^n$}{
    $V_{i-j}' = V_{i-j}^n \left(\sum_{ij} w_{ij} V_{ij}^\text{obs} / y_{ij}^n \right) / \sum_{ij} w_{ij}$\;
    $V_{i-j}^{n+1} = (1-\delta)V_{i-j}^n + \delta V_{i-j}'$\;
  } 
  \If{$||\Delta\mathbf{x}||_2 / ||\mathbf{x}||_2 < \varepsilon$}{
    break\;
  }
 }
\KwResult{Gains $\mathbf{g}$ and visibility solutions $\mathbf{V}$} 
 \caption{\texttt{omnical}} \label{alg:omnical}
\end{algorithm}
There we defined $\mathbf{g}$ as the vector of calibration solutions $g_i$, $\mathbf{V}$ as the vector of visibility solutions $V_{i-j}$, $\mathbf{x}$ as the vector of both, and $\Delta\mathbf{x}$ as the difference between the solutions from one iteration to the next. When we evaluate the gain update, we sum over all antennas. When we evaluate the visibility solution update, we sum over all visibilities with the same baseline separation as $V_{i-j}$. To avoid over-correction on any given step and thus speed up convergence, we damp each update by a factor $\delta$. We generally find that $0.1 \lesssim \delta \lesssim 0.5$ converges fastest. This process repeats until some convergence level $\varepsilon$ in the $L^2$-norm is reached. For HERA we use $\delta = 0.4$ and $\varepsilon = 10^{-10}$.

Ideally, we would weight each visibility by the square of the SNR on each measured visibility. While we could use $V_{ij}^\text{obs}$ as the signal, a less noisy estimate of the quantity is the most recent iteration of $y_{ij}^n$. For the noise, we use the observed visibility autocorrelations, $V_{ii}^\text{obs}$:
\begin{equation}
    \sigma^2_{ij} = \left\langle  \frac{V_{ii} V_{jj}}{\Delta t \Delta \nu} \right\rangle \approx \frac{V_{ii} V_{jj}}{\Delta t \Delta \nu}, \label{eq:noise_var}
\end{equation}
where $\Delta t$ is the integration time and $\Delta \nu$ is the channel bandwidth. The first equality follows from the radiometer equation \citep{thompson_et_al2017}, the later approximation uses the fact that autocorrelations are measured at extremely high signal-to-noise in HERA (usually $\gg$100 in 10.7\,s integrations). In Figure~\ref{fig:chisq_convergence} we show that this proper noise-weighting allows us to converge to $\chi^2/\text{DoF} \approx 1$ quite quickly.

\begin{figure}
    \centering
    \includegraphics[width=.5\textwidth]{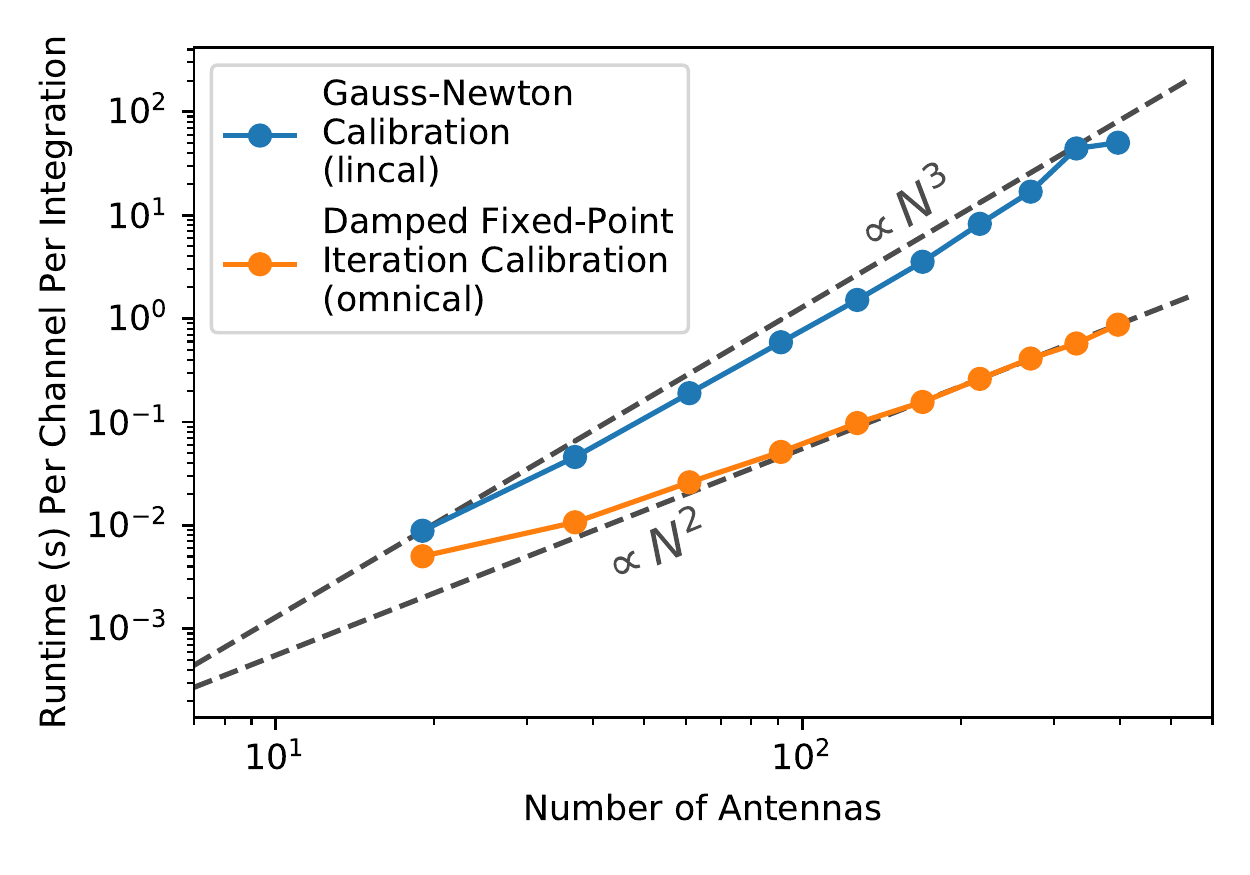}
    \vspace{-10pt}
    \caption{\texttt{omnical}, the damped fixed-point iteration algorithm we describe in Section~\ref{sec:omnical} that avoids matrix inversion converges much faster than the \texttt{lincal} while achieving the same level of precision. Here we show the per-frequency and per-integration runtime for the two algorithms using single-precision floating point variables. The simulation uses increasingly large hexagonal arrays and simulates redundant visibilities using the same technique as the one described in Figure~\ref{fig:chisq_convergence} and Section~\ref{sec:firstcal}.}
    \label{fig:omnical_speed}
\end{figure}

Even though \texttt{omnical} generally takes many more steps to converge than \texttt{lincal}, each step is much faster because it does not involve a matrix inversion. Comparing the total runtime to achieve convergence at the same level of precision, we see in Figure~\ref{fig:omnical_speed} that \texttt{omnical} scales with the number of antennas as $\mathcal{O}(N_\text{ant}^2)$ while \texttt{lincal} scales as $\mathcal{O}(N_\text{ant}^3)$.

There are a number of ways to speed up the algorithm. Since the calibration is independent for each time and frequency, allowing each observation to converge independently is generally a speedup. Being careful to reuse repeated calculations also saves time. In our implementation, checking for convergence is actually a bottleneck; we speed up the algorithm by updating the noise model and checking for convergence every 10 iterations.\footnote{Though for the simulation in Figure~\ref{fig:chisq_convergence}, we did a full update at every iteration so as not to introduce confusing discontinuities in results.} The source code for \texttt{omnical},\footnote{This pure-python implementation is faster than the original C++ implementation of \texttt{omnical} and has fixed the convergence issue identified in Appendix B of \citet{Li_et_al2018}, thus eliminating the need for any final \texttt{lincal} post-processing.} along with \texttt{firstcal}, \texttt{logcal}, and \texttt{lincal} is available freely on GitHub.\footnote{\url{https://github.com/HERA-Team/hera_cal/}}

\subsubsection{Fixing Degeneracies}

Since redundant-baseline calibration is carried out for each frequency and time independently, it will generally be the case that the calibration of different frequencies and times will fall in different parts of the degenerate subspace of the gain and visibility solutions that minimize $\chi^2$. In principle, this is not a problem; absolute calibration is designed to fix this and \citet{kern_et_al2019c} showed that that technique works quite well. However, we have found that absolute calibration (itself an iterative process) converges more quickly and reliably when we start the process by taking out an overall phase slope and an overall delay slope---precisely the degeneracies of \texttt{firstcal}. It is therefore useful, as was argued in \citet{dillon_et_al2017}, to fix the degenerate terms to avoid introducing unnecessary spectral or temporal structure that we must later take out.

Given our absolute calibration strategy and the fact that the degeneracies of \texttt{firstcal} are a subset of the degeneracies of full redundant-baseline calibration, we use our \texttt{firstcal} gains as a degenerate ``reference.'' For the amplitude degeneracy, all \texttt{firstcal} gains have unit amplitude. Therefore, we fix our gains to have an average product over all antenna pairs of 1. For the phase degeneracies, we demand that the average phase and the phase slope (computed by dotting each antenna position into its phase) to be the same after \texttt{firstcal} as after redundant-baseline calibration. We then update the visibility solutions accordingly to keep $\chi^2$ constant.

\subsection{The Normalization of $\chi^2$} \label{sec:chisq_dof}

Before we turn to real HERA data and a thorough investigation of $\chi^2$ as it depends on time, frequency, antenna, and baseline, it is important to understand quantitatively what we expect $\chi^2$ to be for HERA. One might guess that, in a perfectly redundant array, $V_{ij}^\text{obs}$ and $g_i g_j^* V_{i-j}$ differ only by the noise on the observed visibility and thus that expectation value of Equation~\ref{eq:chisq}, $\langle \chi^2 \rangle$, should simply be the number of baselines, $N_\text{bl}$. This would be true if we had an external way to estimate $g_i$ and $V_{i-j}$, but since we do not, we must account statistically for the fitting of noise. After all, a small enough system of equations with enough free variables can always find a solution such that $\chi^2 = 0$. 

\subsubsection{Overall degrees of freedom} \label{sec:overall_dof}

\begin{figure}
    \centering
    \includegraphics[width=.5\textwidth]{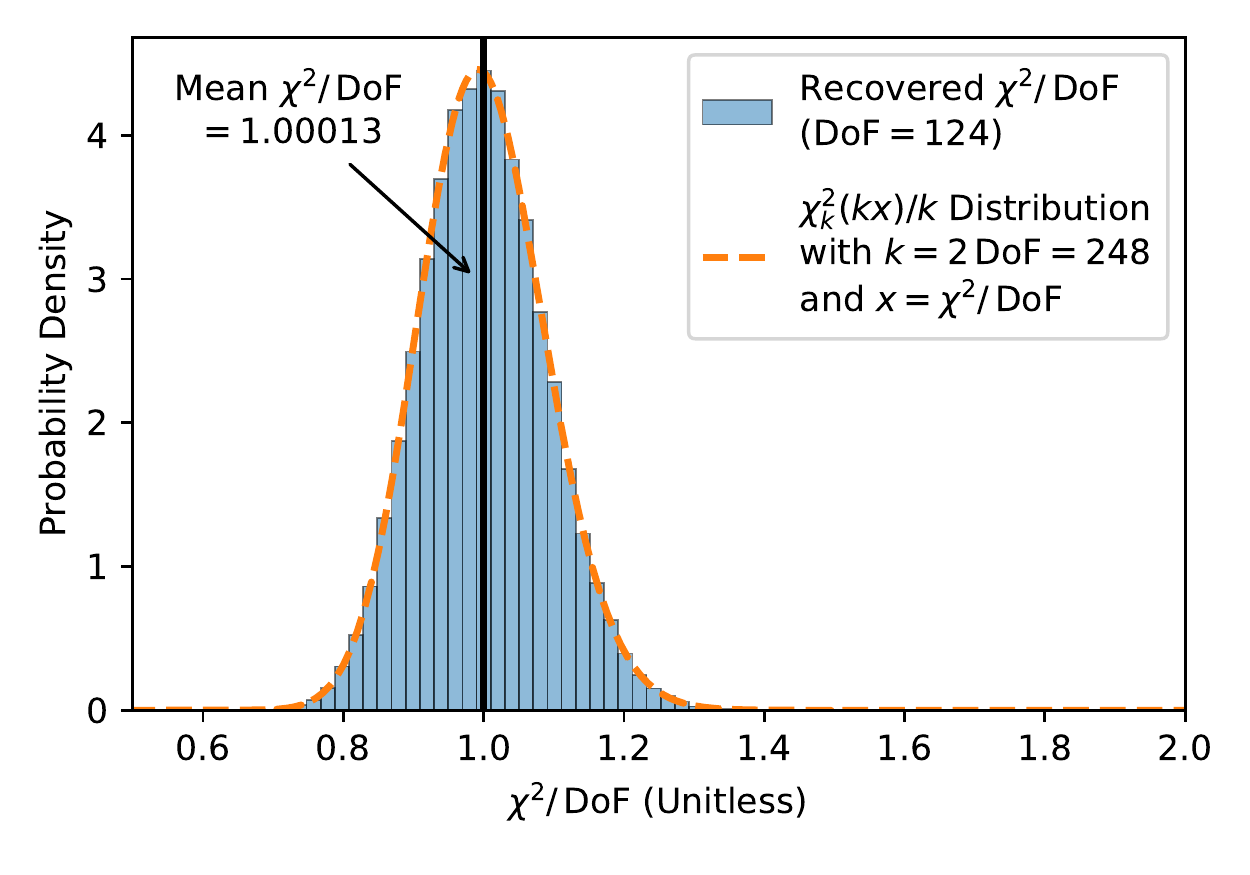}
    \vspace{-10pt}
    \caption{A proper normalization by the number of degrees of freedom in redundant-baseline calibration yields a histogram of simulated $\chi^2 / \text{DoF}$ values consistent with the expected underlying $\chi$-distribution. Here we simulate and calibrate a 19-element hexagonal array, using an analogous technique as laid out in Section~\ref{sec:firstcal} and used for Figure~\ref{fig:chisq_convergence}. Both the mean and variance of this simulated distribution are consistent with our expectations, indicating both good convergence of the algorithm and a correct accounting of the number of degrees of freedom.}
    \label{fig:chisq_dist_sim}
\end{figure}

The actual number of degrees of freedom in single-polarization redundant-baseline calibration is given by
\begin{equation}
    \langle \chi^2 \rangle \equiv \text{DoF} = N_\text{bl} - N_\text{ubl} - N_\text{ant} + 2. \label{eq:dof}
\end{equation}
Here $N_\text{ubl}$ is the number of unique baselines (or equivalently, different $V_{i-j}$ estimated). Intuitively, the number of degrees of freedom is given by the number of measurements minus the number of free parameters solved for in redundant-baseline calibration, i.e.\ the gains and visibilities. However, since each measurement and parameter is complex, the extra two comes from the four real degeneracies, which reduces the number of parameters actually solved for. Equation~\ref{eq:dof} differs from the one that appears in \citet{zheng_et_al2014}, which lacks the two. When the number of baselines is large---in the data discussed in Section~\ref{sec:hera_redundancy} DoF is as large as 533---the error is quite small, which perhaps explains why it was not caught earlier. In Figure~\ref{fig:chisq_dist_sim}, we show the distribution of $\chi^2/\text{DoF}$ using Equation~\ref{eq:dof} for a 19-element hexagonal array simulated analogously to that in Figure~\ref{fig:chisq_convergence} (see Section~\ref{sec:firstcal} for details). 
With 19 elements, we have 171 visibilities and only 30 unique baselines, yielding 124 degrees of freedom. With that normalization, we find that the average $\chi^2/\text{DoF}$ over 102,400 samples (1024 channels and 100 integrations) is 1.00013, consistent with 1.0. Had we used the \citet{zheng_et_al2014} formula, we would have gotten 1.0156.

Furthermore, we see that the distribution of $\chi^2/\text{DoF}$ follows a $\chi^2$-distribution with $k = 2\times\text{DoF}$ degrees of freedom, the factor of two again resulting from the fact that each complex degree of freedom is equivalent to two real degrees of freedom.\footnote{
\label{fn:chisq_dof_dist}This factor of 2 is missing from the discussion in Section~3.1.4 of \citet{zheng_et_al2014}. Likely, this was an oversight that does not affect Figure~11 of that work.
To be precise, if the probability density function of the $\chi^2$-distribution denoted $\chi^2_k(x)$ is given by
$    f_{\chi^2}(x; k) = \frac{1}{2^{k/2}\Gamma(k/2)} x^{k/2 - 1} e^{-x/2}$,
then probability distribution function of $\chi^2/\text{DoF}$ is given by
${f\left(\frac{\chi^2}{\text{DoF}}; DoF\right) = f_{\chi^2}(2\chi^2; 2\times\text{DoF}) / \left(2 \times \text{DoF}\right)}$. 
We choose not to simplify further because it is generally easier to compute this function numerically using a standard library for $f_{\chi^2}(x; k)$, e.g.\ \texttt{scipy.stats.chisq.pdf()}.} 
We see in Figure~\ref{fig:chisq_dist_sim} that this functional form fits the simulated histogram quite well. Given that the expected variance of $\chi^2_k(x)$ is $2k$, then it follows that the expected variance of $\chi^2/\text{DoF}$ should be $\text{DoF}^{-1}$. Indeed, we find that $\text{Var}\left(\chi^2/\text{DoF}\right) \times \text{DoF} = 1.0028$, consistent with 1.0.

\subsubsection{Per-baseline degrees of freedom} \label{sec:per-bl-dof}

While $\chi^2/\text{DoF}$ is a very useful summary statistic for how consistent our observations are with thermal noise, it is generally useful to break apart the sum in Equation~\ref{eq:chisq} to assess redundancy as a function of baseline, unique baseline, or antenna. We know that the sum of all $\langle\chi^2\rangle$ per baseline or per unique baseline group should be the total $\text{DoF}$. Likewise, if we define $\chi^2$ per antenna such that each term in the sum in Equation~\ref{eq:chisq} is assigned to both antennas involved in the visibility, then it follows that the sum of all $\langle\chi^2\rangle$ per antenna should be $2\times\text{DoF}$. It is not true, however, that the degrees of freedom are equally distributed among the baselines or antennas. In the case of baselines, it is trivial to see this; adding a baseline that is not redundant with any other baseline adds one new complex data point and one new complex variable. We expect that for that baseline, we can always find a $V_{i-j}$ such that $V_{ij}^\text{obs} = g_i g_j^* V_{i-j}$ exactly, meaning that $\chi^2$ for this baseline should always be 0.  

After some numerical exploration and educated guesses, we found a method that predicts $\chi^2$ per baseline quite accurately. Specifically, we found that the expectation value of the vector of $\chi^2$ values per baseline, $\pmb{\chi}^2$, is given by
\begin{equation}
    \langle \pmb{\chi}^2 \rangle = \mathbf{1} - \frac{1}{2} \text{Diag}\left[\mathbf{A} \left(\mathbf{A}^\intercal\mathbf{A}\right)^{-1}\mathbf{A}^\intercal + \mathbf{B} \left(\mathbf{B}^\intercal\mathbf{B}\right)^{-1}\mathbf{B}^\intercal  \right] \label{eq:chisq_per_bl}
\end{equation}
where $\mathbf{A}$ and $\mathbf{B}$ are the real and imaginary \texttt{logcal} matrices defined in Equation~\ref{eq:logcal_matrices} and the matrix inversions are actually Moore-Penrose pseudoinverses, since both $\mathbf{A}^\intercal\mathbf{A}$ and $\mathbf{B}^\intercal\mathbf{B}$ are rank-deficient by the number of degeneracies of redundant-baseline calibration (1 for $\mathbf{A}^\intercal\mathbf{A}$, 3 for $\mathbf{B}^\intercal\mathbf{B}$). 

\begin{figure*}
    \centering
    \includegraphics[width=\textwidth]{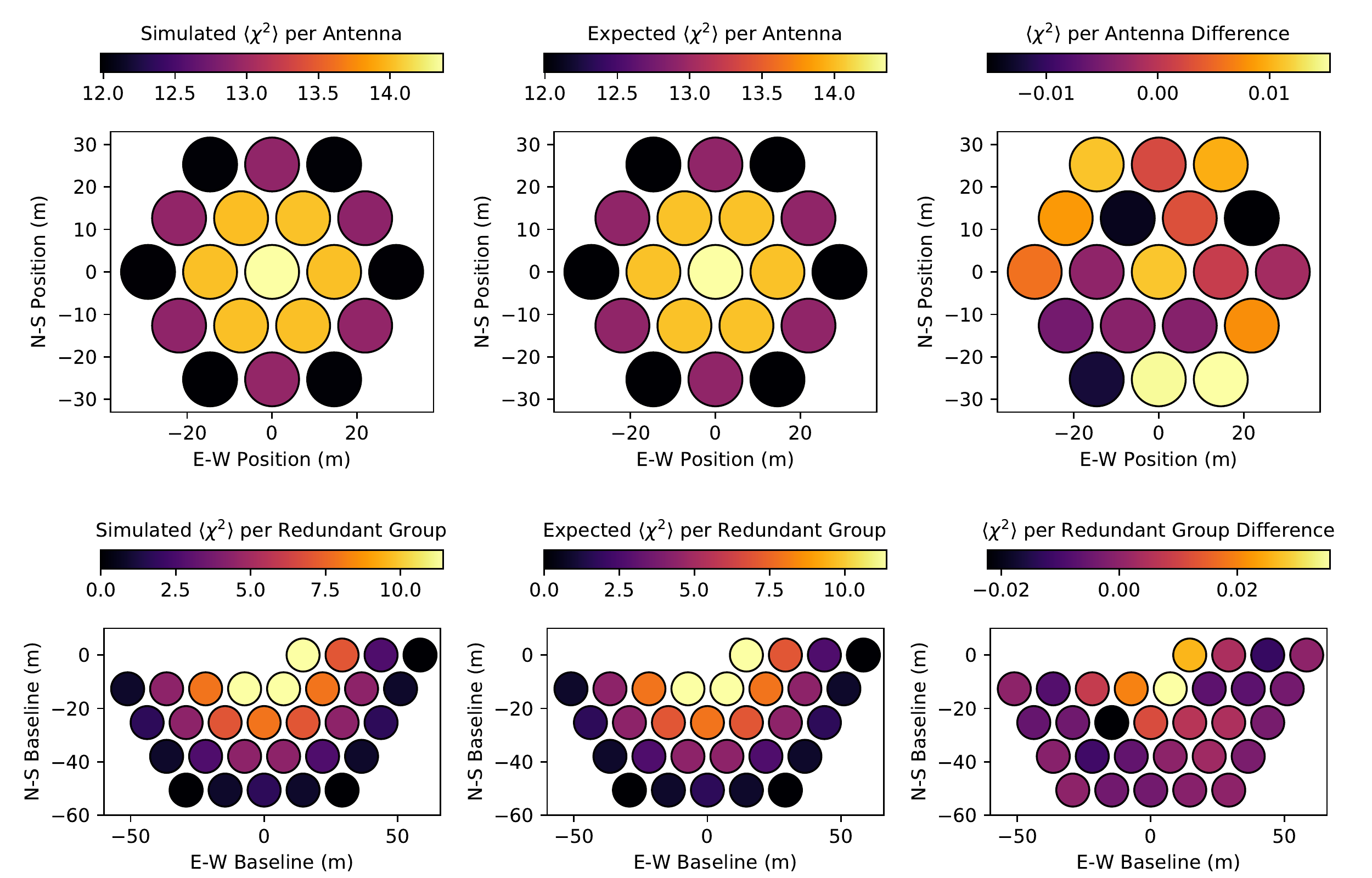}
    \vspace{-5pt}
    \caption{Using Equation~\ref{eq:chisq_per_bl} we can predict how $\chi^2$ should break down by baseline---and therefore by antenna or by unique baseline. In the first row, we show our prediction for $\chi^2$ per antenna, calculated by assigning each term in the $\chi^2$ sum in Equation~\ref{eq:chisq} to both antennas involved. In the second row, we instead break up the sum into unique baseline groups. Using the same simulation as in Figure~\ref{fig:chisq_dist_sim}, we show  in both cases that the average simulated $\chi^2$ (left column) matches the predicted value (middle column) to $\sim$1\% accuracy (right column).}
    \label{fig:chisq_per_red_and_ant_sim}
\end{figure*}

While in Figure~\ref{fig:chisq_per_red_and_ant_sim} we show numerically that Equation~\ref{eq:chisq_per_bl} works quite well, we have been unable to prove it analytically. 
To numerical precision, the sum of $\langle\pmb{\chi}^2\rangle$ over all baselines matches the value in Equation~\ref{eq:dof}. However the match is not perfect, perhaps due to the biases in \texttt{logcal} or an insufficient accounting for the signal-to-noise ratio on each baseline.

We can, however, get some intuition for why this might work. The matrix $\mathbf{A} \left(\mathbf{A}^\intercal\mathbf{A}\right)^{-1}\mathbf{A}^\intercal$ is often referred to as the \emph{data resolution matrix} in the geophysics literature \citep{menke_1989} and is essentially the extent to which the data postdicts itself. That is to say that if we have a system of linear equations $\langle\mathbf{d}\rangle = \mathbf{Mx}$ with equally-weighted data $\mathbf{d}$ and parameters $\mathbf{x}$, then the postdicted set of data using the parameters inferred from the measured data is 
\begin{equation}
    \mathbf{d}^\text{postdicted} = \mathbf{Mx}^\text{inferred} =  \mathbf{M}\left(\mathbf{M}^\intercal\mathbf{M}\right)^{-1}\mathbf{M}^\intercal \mathbf{d}^\text{measured}.
\end{equation}
When the data resolution matrix is the identity, the parameters contain all the information in the data. In our case, that would mean that every baseline's $\chi^2$ should be 0. Therefore, the more  that each piece of data is predicted by other data, the more off-diagonal the data resolution matrix is, the less the noise is fit by the parameters and thus the more degrees of freedom there are.

%%%%%%%%%%%%%%%%%%%%%%%%%%%%%%%%%%%%%%%%%%%%%%%%%%%%%%%%%%%%%%%%%%%%%%%%%%%%%%%%%%%%%%%%%%%%%%%%%%%%%%%%%

\section{Assessing HERA Redundancy} \label{sec:hera_redundancy}

With both algorithms for performing redundant-baseline calibration and mathematical structures for assessing its success, we can now turn to an assessment of the redundancy of Phase I HERA data. In Section~\ref{sec:observations}, we will present the observations we analyzed. In Section~\ref{sec:malfunctioning}, we will explain how redundant-baseline calibration informed our data quality assessment and selection process by looking for outliers in $\chi^2$ per antenna. Next, we we will examine how $\chi^2$ breaks down as a function of time and frequency (Section~\ref{sec:overall_chisq}) and by antenna and redundant-baseline group (Section~\ref{sec:chisq_structure}). Finally, in Section~\ref{sec:fractional_nonred}, we will take a different approach to assessing non-redundancy and offer an answer to the more intuitive but less mathematically well-defined question: \emph{how redundant is HERA?}

\subsection{Observations} \label{sec:observations}

The data in this section come from 18 nights of observation with HERA between December 10, 2017 and December 28, 2017, corresponding to all Julian dates between 2458098 and 2458116 except 2458100 when the correlator was malfunctioning. HERA is located in the Karoo Radio Astronomy Reserve at $-30.7215^\circ$ latitude, $21.4283^\circ$ longitude. As a zenith-pointing, drift-scan array, HERA observations measure a roughly 10$^\circ$ stripe (the full-width at half-maximum of the primary beam) at 150\,MHz, centered at $-30.7215^\circ$ declination. Since HERA is sky-noise dominated, observations where the sun is above the horizon are flagged, since they are both noisier and less redundant than nighttime observations. This means that the observations span a range of local sidereal times of 0.164--11.566\,hours, corresponding to zenith right ascensions in the range of $2.46^\circ$--$173.49^\circ$ .

\begin{figure}
    \centering
    \includegraphics[width=.5\textwidth]{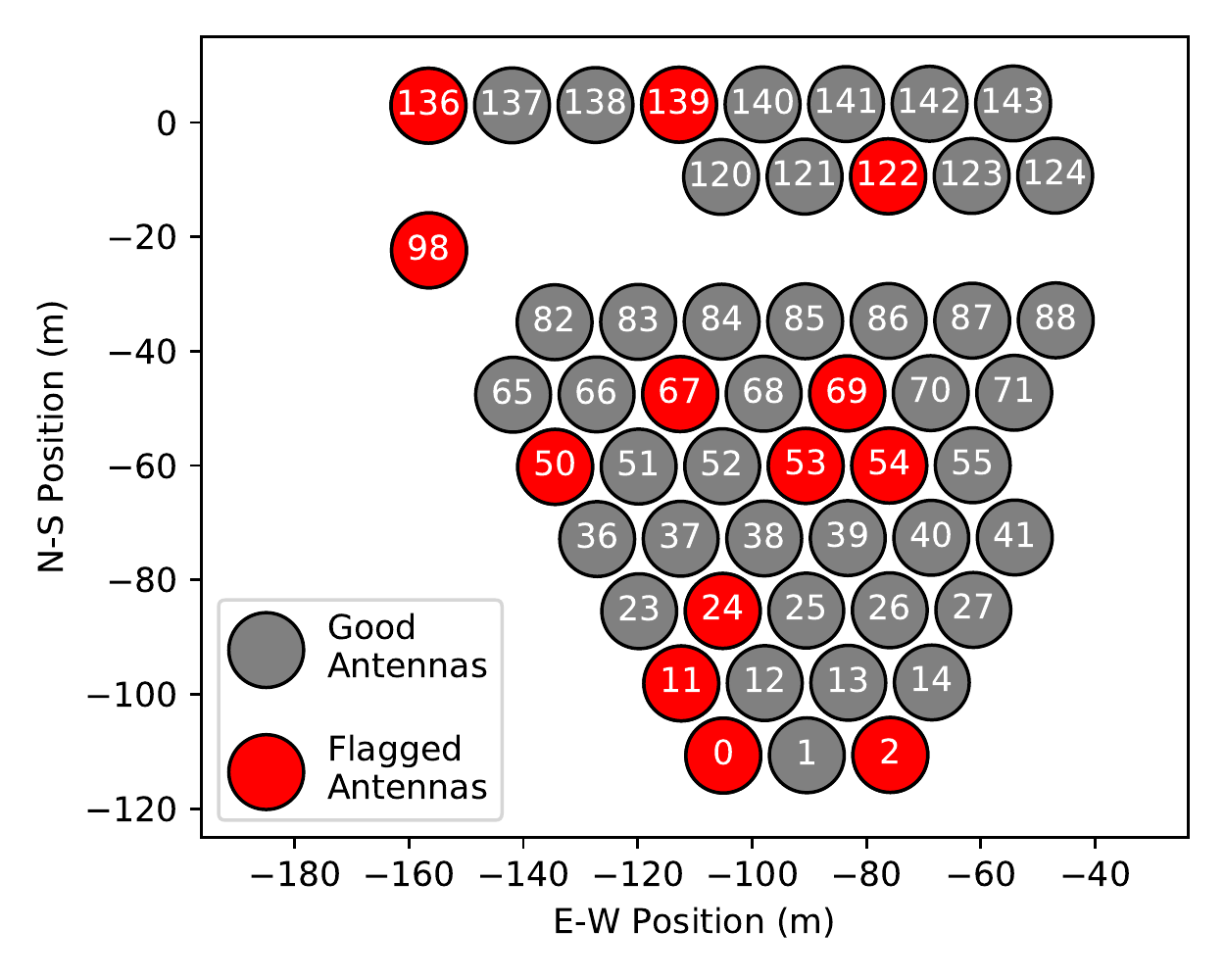}
    \vspace{-10pt}
    \caption{Layout of HERA antennas between Julian dates 2458098 and 2458116. These 14-m diameter dishes are on a hexagonal grid with a separation of 14.6\,m and will eventually be part of the Southwest sector of the split HERA core when HERA-350 is complete. Antennas flagged for non-redundancy or otherwise suspected of malfunctioning (see Section~\ref{sec:malfunctioning}) are noted in red and excluded from the final redundant-baseline calibration examined in this work.}
    \label{fig:array_layout}
\end{figure}

During Phase I, HERA observed from 100--200\,MHz in 1,024 frequency channels, though the upper and lower $\sim$50 channels were flagged because the feed design and the band-limiting filters eliminated most of array's sensitivity to those frequencies. 
Visibilities were measured with 10.7\,s integrations. At that time the array consisted of 52 antennas, of which we eventually threw out 13 due to non-redundancy or other issues (see Section~\ref{sec:malfunctioning}). We show the configuration of the array and the numbering of antennas, including the flagged antennas, in Figure~\ref{fig:array_layout}. 
All of these antennas are part of Southwest sector of the split-core of the eventual HERA-350 design \citep{deboer_et_al2017, dillon_parsons2016} and thus on the same hexagonal grid. 

For most of the redundancy metrics we assess in Sections \ref{sec:overall_chisq}--\ref{sec:fractional_nonred}, we exclude frequencies and times identified as possibly containing RFI. The process for identifying RFI looks at both the data and a variety of reduced data products, including the \texttt{omnical} gains and visibilities, for outliers relative to neighboring times and frequencies. Assuming RFI events are usually compact in time, frequency, or both, the technique then looks for marginal outliers neighboring strong outliers and grows the flagged region accordingly \citep{Kerrigan_et_al2019}. The flags are then harmonized to a single function of frequency and time for all antennas and visibilities, flagging completely the channels or integrations that show low-level contamination in the context of the full data set. A full description of this work will appear the discussion of the forthcoming HERA Phase I power spectrum upper-limits paper.

\subsection{Identification of Malfunctioning Antennas With High-$\chi^2$} \label{sec:malfunctioning}

After each night of observation, HERA's ``real-time'' pipeline (RTP) identified antennas with particularly low power and flagged them \citep{Ali:thesis}. During the period analyzed here, Antennas 0 and 50 were flagged. To be conservative, antennas were flagged even if only one of the two antenna polarizations was malfunctioning.

Redundant-baseline calibration---in this case performed well after the observations were taken\footnote{Though in future HERA observations, we plan to use these algorithms to perform redundant-baseline calibration within 24 hours of taking the data to provide actionable information on the health of the array to the site team.}---gives us another tool with which to assess the health of the array: $\chi^2$ per antenna. Armed with a proper normalization of the expected degrees of freedom in $\chi^2$ per antenna derived from Equation~\ref{eq:chisq_per_bl}, we can look for outliers. 

\begin{figure*}
    \centering
    \includegraphics[width=\textwidth]{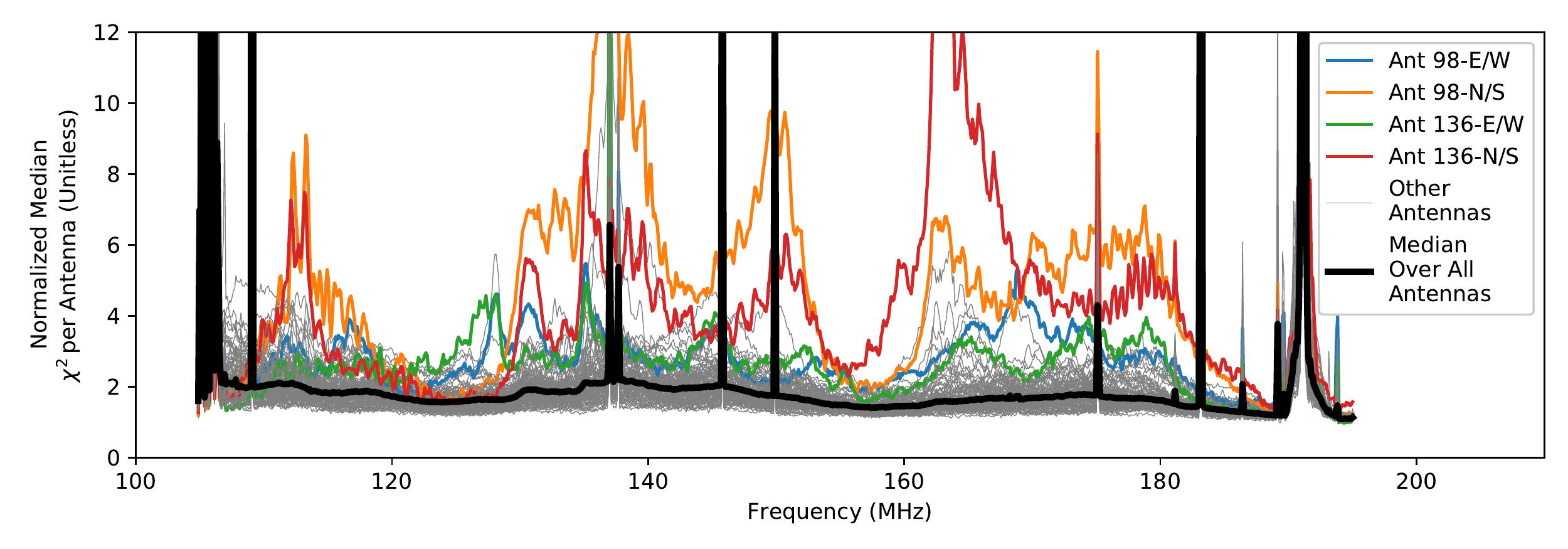}
    \vspace{-10pt}
    \caption{Here we show an example of our observed, normalized $\chi^2$ per antenna, taking the median over time in single file (60 10.7\,s  integrations). The data were taken from a field at $\sim$10\,hours of LST on 2458114 have had only Antennas 0 and 50 removed. The resulting data show that Antennas 98 and 136 remain clear outliers from the others, meriting flagging for extreme non-redundancy. This process is then repeated as eliminating outliers makes less clear outliers stand out better.}
    \label{fig:example_cspa_outliers}
\end{figure*}

In Figure~\ref{fig:example_cspa_outliers} we show the result of our calibration of a single 60-integration file on 2458114 centered on LST $\approx 10$\,hours after removing only antennas 0 and 50.
Two antennas, 136 and 98, stand out right away---especially the North/South-oriented polarizations, though unsurprisingly their East/West-polarizations also appear to be outliers. The high per-antenna $\chi^2$ means that the visibilities these two antennas participate in are particularly discrepant with other visibilities in the same redundant groups. The fact that these antennas are on the edge of the array and thus have different distribution of baselines that they participate in should be taken into account by our DoF normalization (see Figure~\ref{fig:chisq_per_red_and_ant_sim}). While it is difficult to know for certain, perhaps the telephone poles from which the feeds are suspended over the dishes were sometimes less well-balanced at the edge of the array (normally they support three feeds each at $120^\circ$ angles), moving the feed off-center or out of focus and thus creating non-redundancy.

Instead of visually inspecting every piece of data for outliers in $\chi^2$, we quantify the ``outlierness'' of an antenna using its modified $z$-score, defined as
\begin{equation}
	z(x) \equiv 0.6745 \left(\frac{x - \text{Median}(x)}{\text{MAD}(x)}\right)
\end{equation}
where the denominator of the right-hand side is the median absolute deviation (MAD), the median absolute value of the difference between a data point and the median of the data points. The normalization of 0.6745 ensures that, with Gaussian-distributed data, a $1\sigma$ outlier in the standard $z$-score (which uses means and standard deviations) would also be a $1\sigma$ outlier in the modified $z$-score. 

In our case, we first compute the median value of each antenna's $\chi^2$ over all frequencies and times in a single file. This avoids giving undue influence to observations with very high $\chi^2$ due to RFI (seen as spikes in Figure~\ref{fig:example_cspa_outliers}). Then we take these single median $\chi^2$ values for each antenna and compute the median and MAD over all antennas. This gives us a modified $z$-score, which is less sensitive to extreme outliers than a standard $z$-score. 

That said, outliers in redundancy cannot always be so cleanly identified with just a single round of calibration. While redundant-baseline calibration has the virtue of insulating the calibration at a given frequency and time from bad frequencies and times (e.g.\ when there is a strong RFI event), it necessarily cannot isolate bad antennas from the rest of the gain and visibility solutions. Every antenna is involved in baselines with the worst antennas, which means that every antenna's $\chi^2$ will be somewhat elevated. Our strategy is thus to remove all antennas that are $4\sigma$ outliers and then recalibrate, lowering the median over all antennas and exposing new outliers. We repeat until no new outliers appear.

\begin{figure*}
    \centering
    \includegraphics[width=1\textwidth]{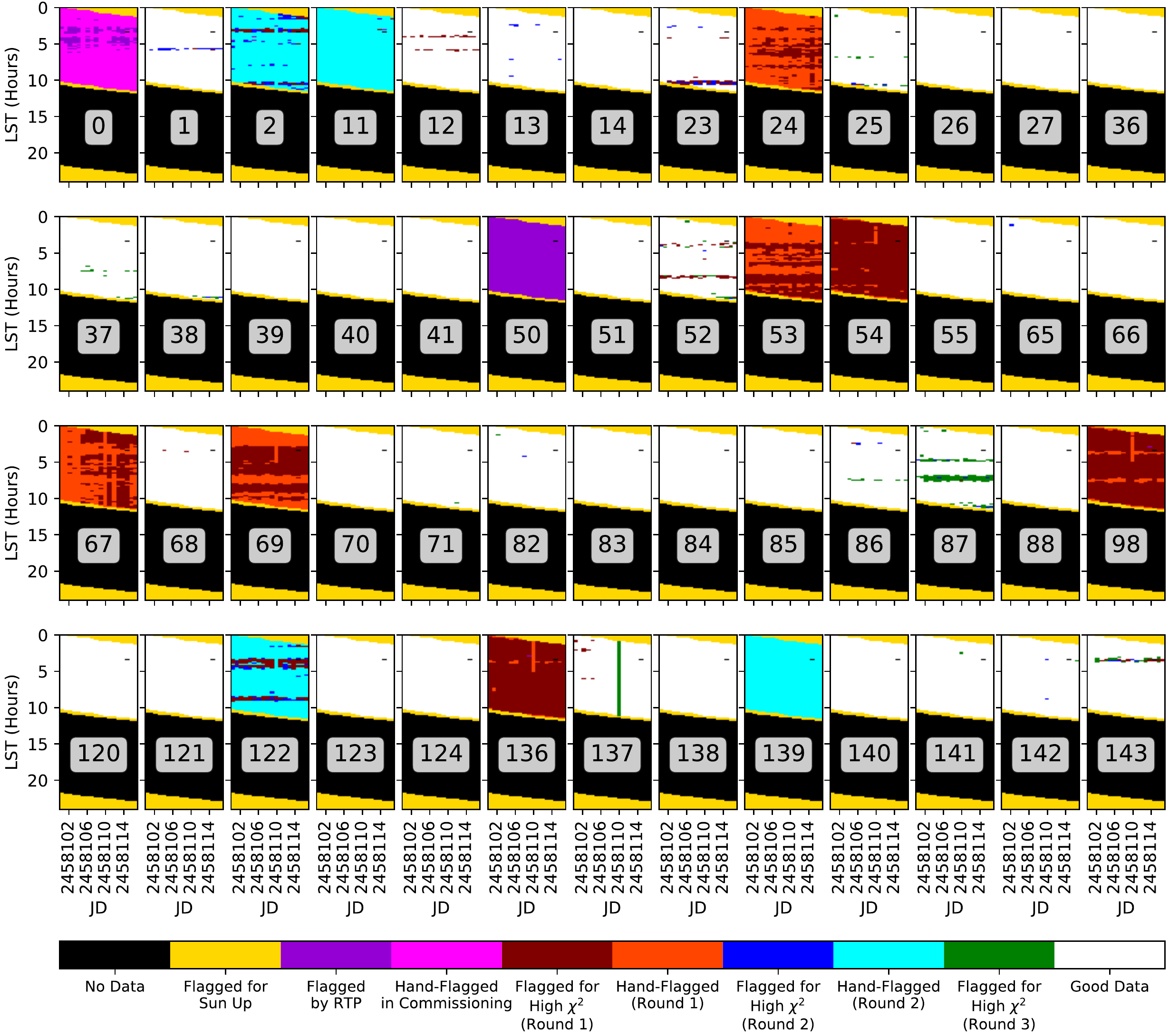}
    \vspace{-10pt}
    \caption{Here we show our per-antenna data quality assessment and flagging as a function of night and LST. Each pixel shows whether an antenna was completely flagged for a file and at what stage in the process. After throwing out daytime data (yellow) and bad antennas caught by the RTP and the commissioning team (purple and magenta), we proceed through three rounds of redundant-baseline calibration of the entire data set, iteratively flagging $4\sigma$ outliers in the modified-$z$ score of their per-antenna $\chi^2$ (maroon, dark blue, green). After the first two rounds of calibration and flagging, we hand-flag antennas (orange, light blue) that were frequently identified as outliers or were externally removed in subsequent processing \citep{kern_et_al2019c}. This gives us the data set we analyze in the rest of this work and from which we plan to produce the first HERA EoR power spectrum limits. (The one piece of data missing from every antenna during the night of 2458114 is due to a correlator restart.)}
    \label{fig:antenna_flagging}
\end{figure*}

That is not quite the end of the story. When we calibrate the entire dataset in this way, we find that our method for identifying bad antennas does not always produce consistent results within each night or across nights. In Figure~\ref{fig:antenna_flagging} we show the results of our outlier detection across the entire dataset, condensing each file to a single pixel.
After removing data when the sun was above the horizon (yellow) and antennas flagged by the RTP (purple) or declared suspect by the HERA commissioning team based on the RTP results (magenta), the result of this outlier detection algorithm is shown in maroon. Antennas 54, 98, are 136 are very-consistently identified as bad, but not quite always. Other antennas are flagged inconsistently with flagging patterns that often repeat night-to-night as a function of LST. Clearly, the position of sources affects the level of observed non-redundancy because different antennas are more or less redundant with the bulk of the array along different lines of sight, i.e.\ their primary beams differ. We will return to this observation in Section~\ref{sec:overall_chisq} and assess some of its consequences in Section~\ref{sec:temporal}. 

Given our goal to conservatively select the best, most reliable data, it is prudent to assume that an antenna that is a $4\sigma$ outlier at many LSTs and over many nights is probably a not-quite-$4\sigma$ outlier the rest of the time. We therefore choose to flag Antennas 24, 53, 54, 67, 69, 98, and 136 for the entire data set (orange). We repeat this process one more time, looking for new outliers (dark blue) once all the most consistently bad antennas are removed consistently. On the next round of hand-flagging (light blue), we remove antennas 2 and 122. We also remove antennas 11 and 139 which did not show any substantial non-redundancy but were hand-flagged during absolute calibration \citep{kern_et_al2019c}. We then perform one final round of redundant-baseline calibration with all of those antennas removed and continue to remove the occasional $4\sigma$ outlier (green), but perform no further whole-antenna flagging. Ideally, with redundant-baseline calibration operating in near-real-time, bad antennas can be identified on a nightly basis and removed from future calibration until they are fixed.

\subsection{Overall $\chi^2$ Results} \label{sec:overall_chisq}

With our clear theoretical understanding of the behavior of $\chi^2$ in an ideal array and with our selection of high-quality data, we are now prepared to compare the two. Though $\chi^2$ is a reduced statistic, it is still calculated for every frequency, time, polarization, and night. In this section we examine a few different ways to slice these data and begin to interpret the results.

\begin{figure}
    \centering
    \includegraphics[width=.5\textwidth]{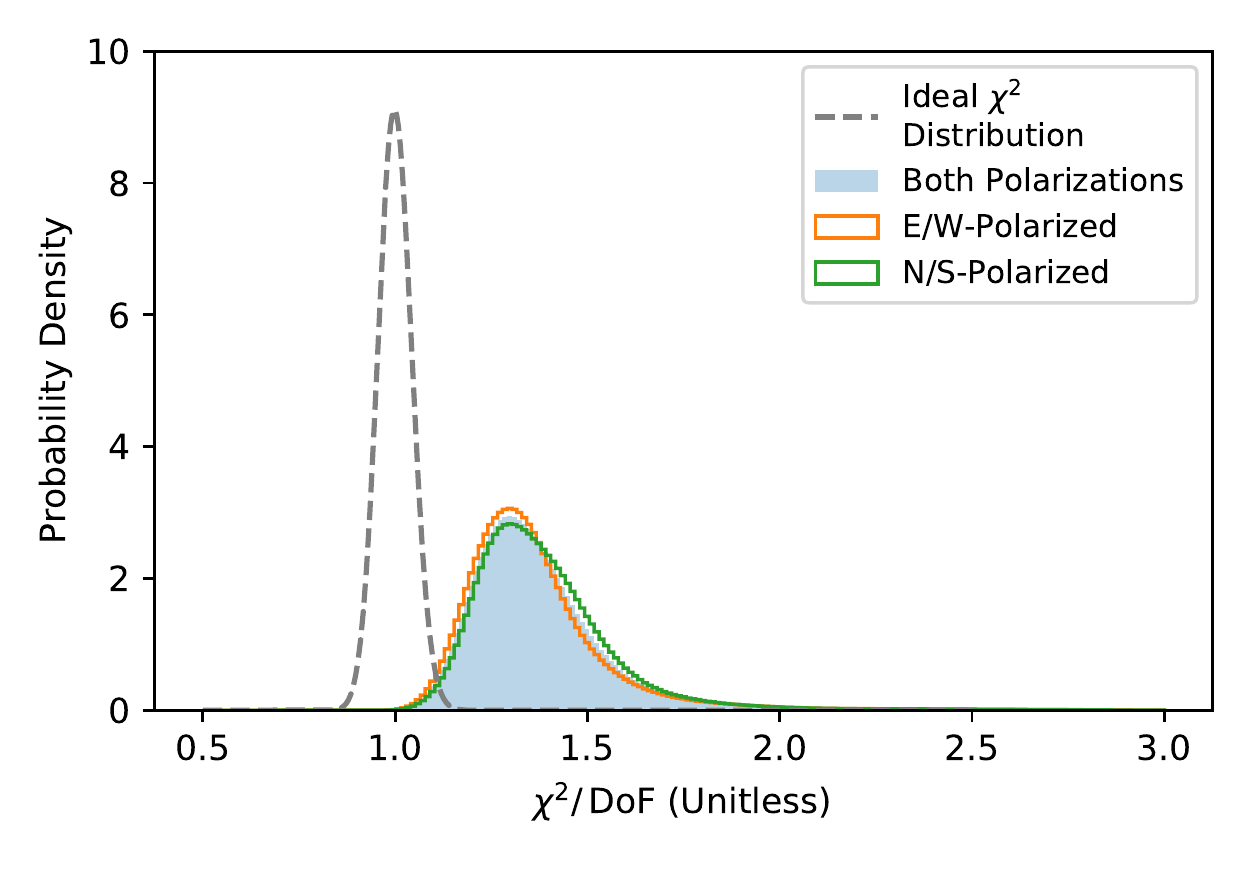} \\ \vspace{-10pt}
    \includegraphics[width=.5\textwidth]{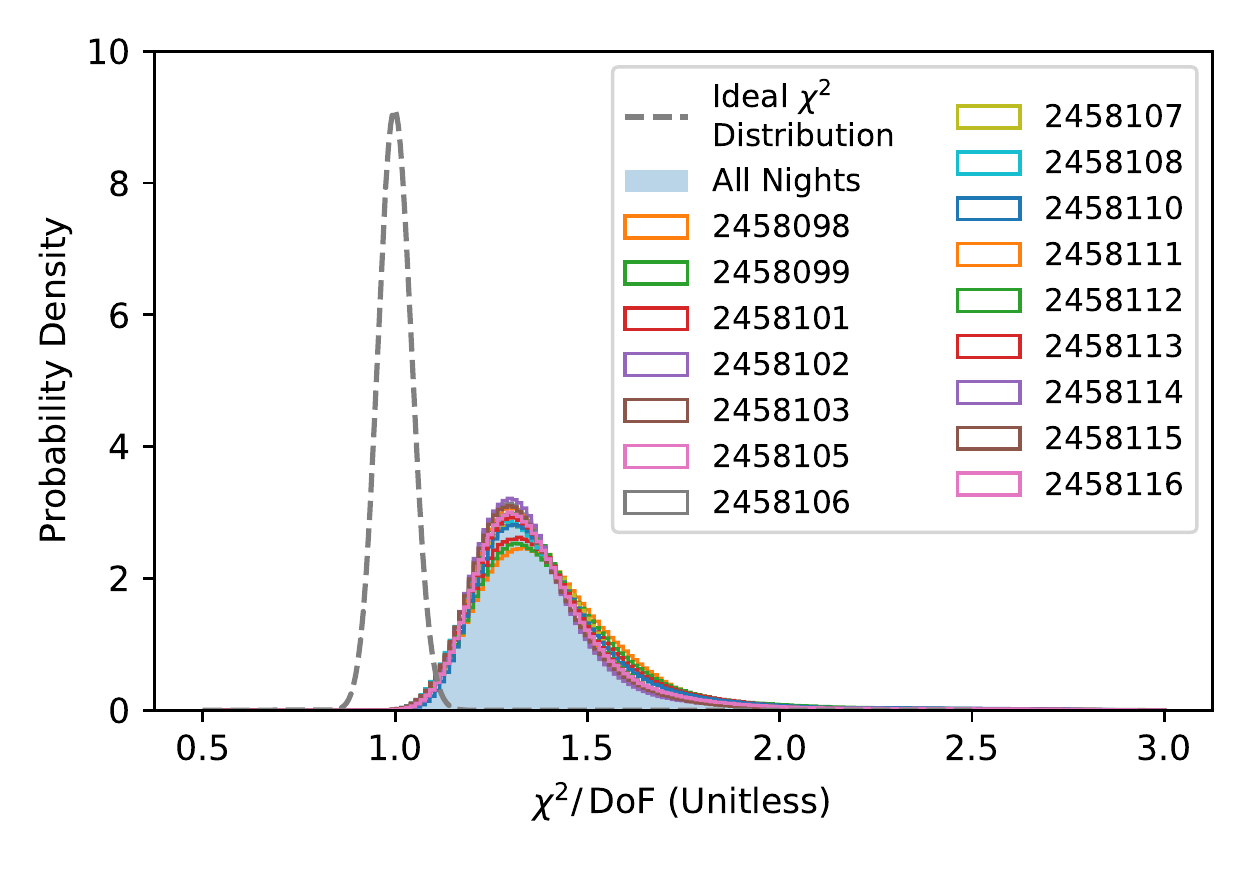} \\ \vspace{-10pt}
    \includegraphics[width=.5\textwidth]{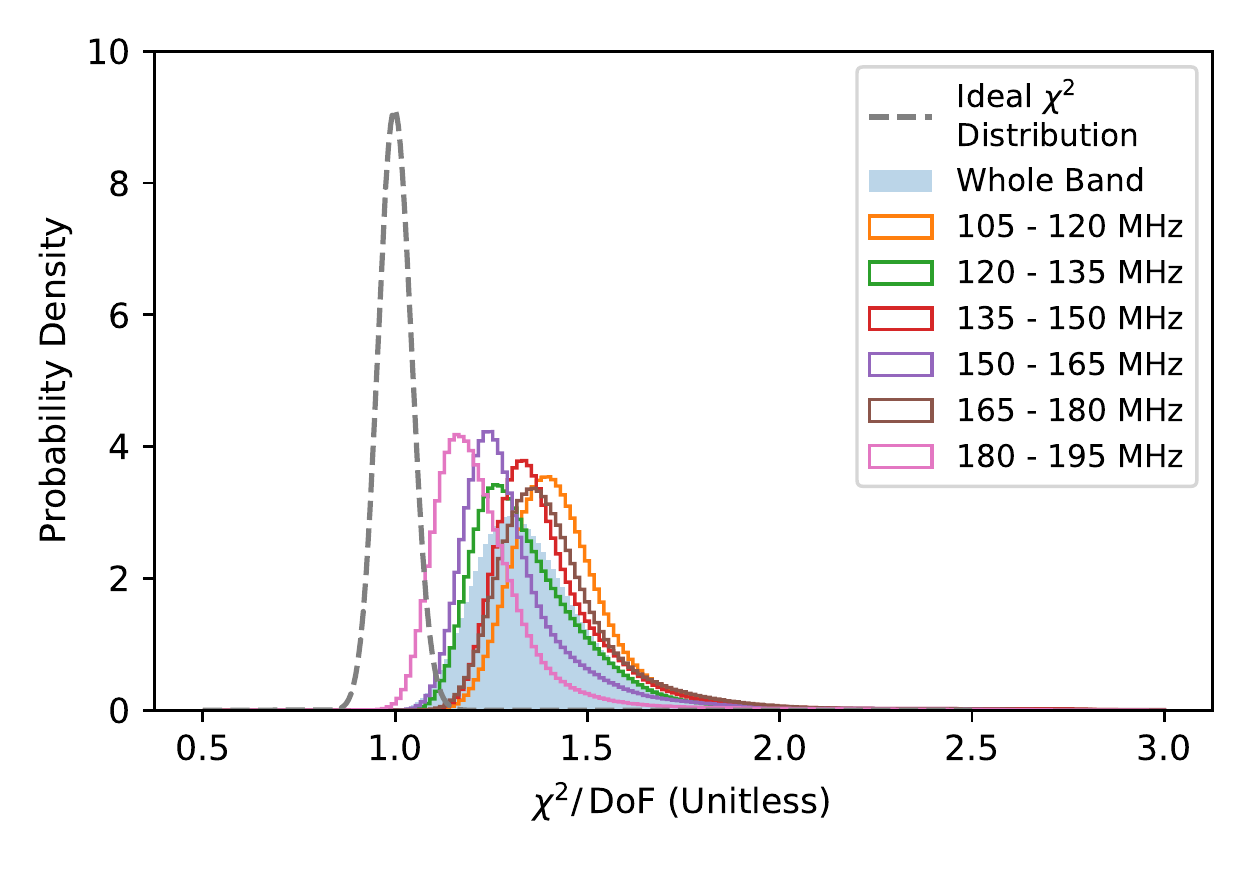}
    \vspace{-10pt}
    \caption{Here we show the distribution of $\chi^2 / \text{DoF}$ over the entire data set. We compare this to an ideal $\chi^2$-distribution (see Footnote \ref{fn:chisq_dof_dist}) with $\text{DoF} \approx 520$, the mean over observations with different antenna flagging.  In contrast to our simulations where the distribution matched the expected one (Figure~\ref{fig:chisq_dist_sim}), here we see clear evidence for the hypothesis that the data cannot be completely described by unique visibility baseline group and one complex gain per antenna---the model for which we are computing $\chi^2$ in Equation~\ref{eq:chisq}. In other words, we see evidence for non-redundancy. This appears to be consistent for both polarizations (top panel) and from night to night (middle panel). However, we do see clear evidence for different distributions when the histogram is broken out into rough frequency bands, indicating varying and non-monotonic levels of non-redundancy as a function of frequency. This effect is seem more clearly in Figure~\ref{fig:chisq_waterfall} and the second panel of Figure~\ref{fig:chisq_medians}.}
    \label{fig:chisq_hists}
\end{figure}

We start by looking at the observed probability density function of $\chi^2$, with an eye towards replicating Figure~\ref{fig:chisq_dist_sim} with real data. We see in Figure~\ref{fig:chisq_hists} our first clear indication of HERA's non-redundancy. 
We plot the distribution of $\chi^2 / \text{DoF}$ separated out along several axes. We compare this to an ideal $\chi^2$-distribution in Footnote \ref{fn:chisq_dof_dist} where we used $\text{DoF} \approx 520$, the mean number of degrees of freedom over our all observations. This is a bit lower than the 533 DoF we would get from only the antenna flags in Figure~\ref{fig:array_layout}. Since each individual $\chi$ measurement is normalized by the correct DoF, the effect of varying per-observation DoF is minimal. 

In most cases, we find that  $\chi^2 / \text{DoF}$ peaks between 1.3 and 1.4, meaning that HERA exhibits persistent non-redundancy $\sim$20\% larger than the thermal noise level. Despite that, the data are extremely inconsistent with the null hypothesis that non-redundancy in visibilities is attributable to pure noise, as the histograms in Figure~\ref{fig:chisq_hists} make clear. Overall, we find a mean value of $\chi^2 / \text{DoF}$ of 1.389. This is somewhat difficult to compare to previous results which have different noise levels normalizing $\chi^2$. In \citet{ali_et_al2015}, PAPER reported a mean $\chi^2 / \text{DoF}$ of 1.9. While PAPER's elements had significantly less collecting area and thus sensitivity, it had substantially larger frequency and time bins (493\,kHz 42.9\,s, compared to 97.7\,kHz and 10.7\,s with HERA). MITEoR reported a mean $\chi^2 / \text{DoF}$ of 1.05, however its integration time (5.37\,s) and frequency resolution (49\,kHz) are both roughly half HERA's and its elements were even smaller. The fact that the instruments have different fields of view and, in MITEoR's case, a northern latitude, further complicate the comparison. Motivated in part by this difficulty, we introduce a relative non-redundancy metric in Section~\ref{sec:fractional_nonred} that is ideally independent of the noise level. We do not believe the elevated $\chi^2 / \text{DoF}$ is attributable to poor convergence due to low visibilities SNR; as \citet{Gorthi_et_al2020} showed in a simulation of a perfectly redundant array, redundant-baseline calibration of an array this size can achieve $\chi^2 / \text{DoF} \approx 1$ even when the visibility SNR is $\lesssim$1.

\begin{figure}
    \centering
    \includegraphics[width=.5\textwidth]{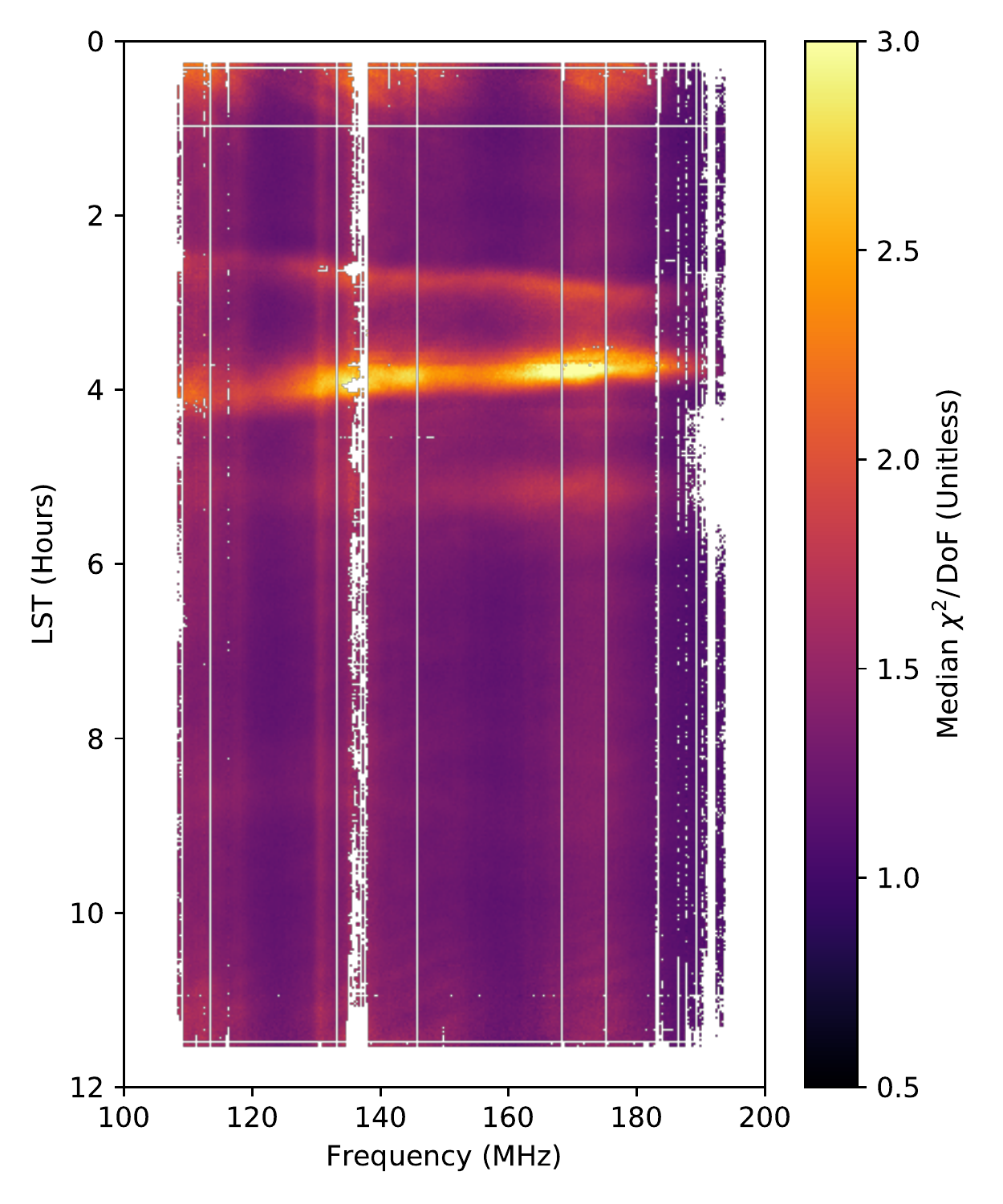} 
    \vspace{-10pt}
    \caption{The median unflagged value of $\chi^2 / \text{DoF}$ over night and polarization reveals complex spectral and temporal structure in this metric of non-redundancy. While close to the ideal value of 1.0 in places, significant deviations are apparent at certain LSTs, especially those associated with bright point sources (see Figure~\ref{fig:chisq_medians}). }
    \label{fig:chisq_waterfall}
\end{figure}

\begin{figure}
    \centering
    \includegraphics[width=.5\textwidth]{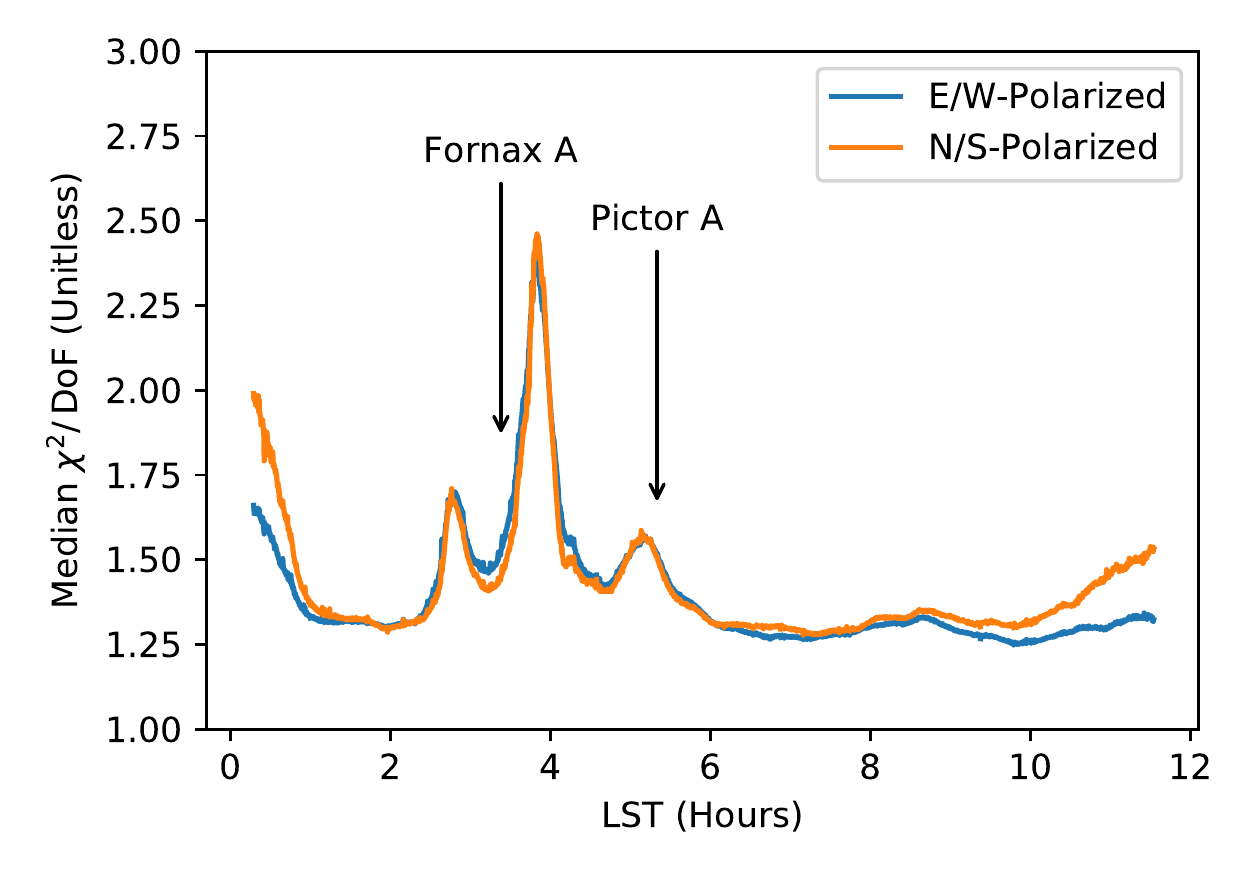} \\ 
    \vspace{-10pt}
    \includegraphics[width=.5\textwidth]{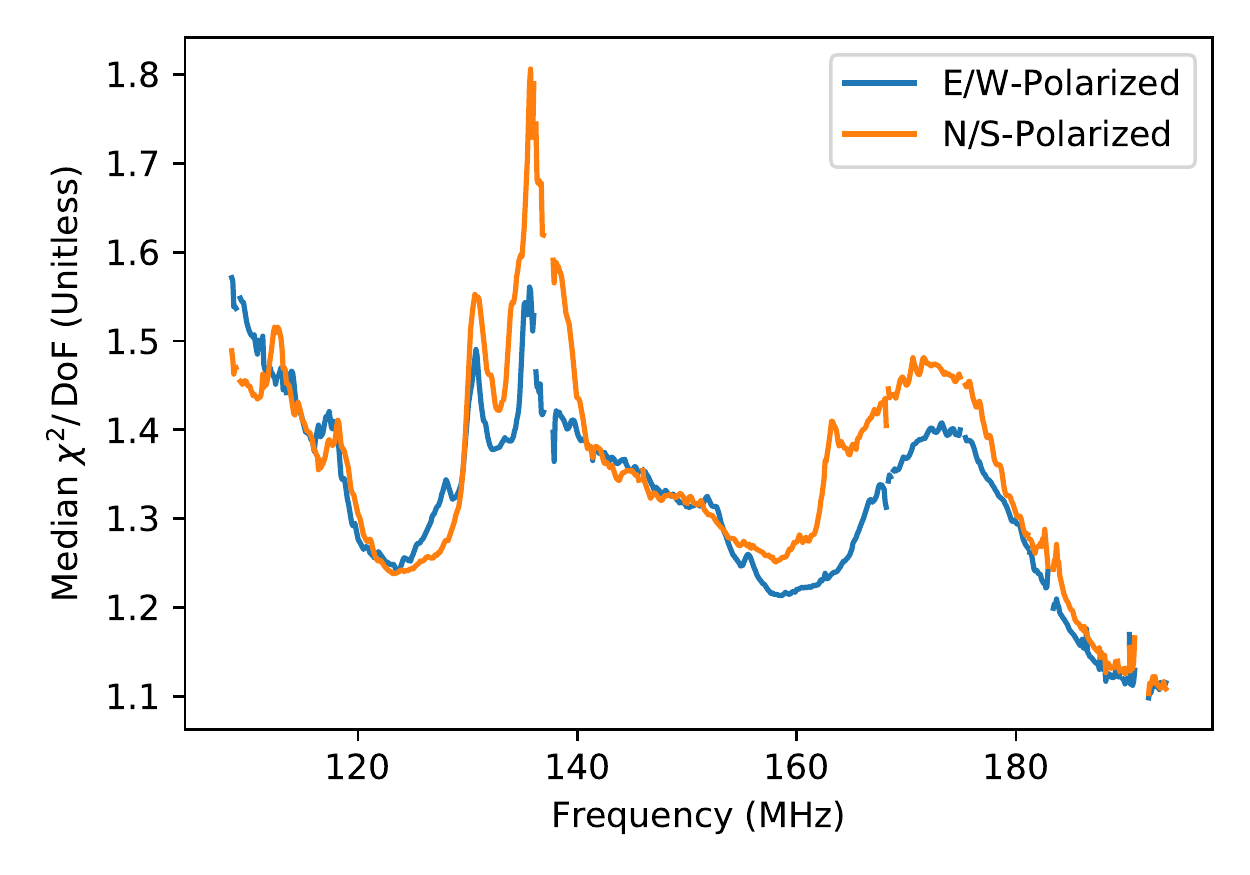} \\ 
    \vspace{-10pt}
    \caption{Collapsing Figure~\ref{fig:chisq_waterfall} along both the frequency and time axis (while breaking it up into two polarizations) reveals significant structure in $\chi^2$, likely due to the primary beam. We attribute the temporal structure to bright point sources moving through the main lobe of the primary beam---lowering $\chi^2$ by making gain and beam variations degenerate---or through the sidelobes---highlighting the antenna-to-antenna variation. We explore this explanation more in Section~\ref{sec:temporal}. The frequency structure appears correlated with beam directivity, though the underlying cause of that correlation is less well understood.}
    \label{fig:chisq_medians}
\end{figure}

The observed $\chi^2 / \text{DoF}$ level appears consistent both for the East/West- and North/South-oriented polarizations and from night to night.\footnote{2458104 and 2458109 are excluded from the middle panel of Figure~\ref{fig:chisq_hists} because they both are flagged due to heavy RFI early in the night. That flagging includes Fornax A's transit, making their $\chi^2$ distributions appear artificially low compared to other nights (see Figures~\ref{fig:chisq_waterfall} and \ref{fig:chisq_medians}).} However, it does exhibit an interesting and clearly non-monotonic dependence on frequency. To understand that structure better, it is easier to look at the median of these distributions (the median is taken to avoid any low-level RFI) as a function of frequency, LST, or both, marginalizing over night and polarization which seem to have little effect. The results are shown in Figures~\ref{fig:chisq_waterfall} and \ref{fig:chisq_medians}. 
Figure~\ref{fig:chisq_waterfall} shows clear evidence of temporal and spectral structure in $\chi^2 / \text{DoF}$, both of which are summarized in Figure~\ref{fig:chisq_medians}. 

The temporal structure of  $\chi^2 / \text{DoF}$, seen both in Figures~\ref{fig:chisq_waterfall} and \ref{fig:chisq_medians} is the easier of the two to understand. By far the largest excess is clearly associated with the transit of Fornax A through the beam. Fornax A, which is at 3.378\,hours of right ascension and $-37.21^\circ$ declination, transits relatively deep into HERA's primary beam, but at 750\,Jy at 154\,MHz \citep{McKinley_et_al2015} it is substantially brighter than anything else in the field.
At roughly 1$^\circ$ across, Fornax A is effectively a very bright point source for HERA Phase I since it is slightly smaller than the synthesized beam. When it transits through the main beam, it is the dominant source in the field.

In the extreme case when the sky is a single point source, antenna-to-antenna beam variations are reduced to a single pierce-point as a function of time and frequency, meaning that they can be wholly subsumed into temporal and spectral gain variations without raising $\chi^2$. This is consistent with the finding that redundancy of closure phases also improves when Fornax A transits \citep{hera_memo_48}. When Fornax dominates, beam and gain variations become less distinguishable and  $\chi^2 / \text{DoF}$ becomes a poorer metric of non-redundancy. By contrast, when Fornax A is in a sidelobe, its apparent flux density likely varies more from antenna to antenna because sidelobes vary comparatively more than the main lobe, due to the increased relative impact of cross-coupling and other non-idealities \citep{Fagnoni_et_al2019a}. As long as other sources of comparable apparent brightness are elsewhere in the beam, that variation along the line of sight to Fornax A cannot be absorbed into the gains. This explains the dual-peaked structure around Fornax A in Figure~\ref{fig:chisq_medians} and the peak near the transit of Pictor A, which at a declination of $-45.78^\circ$ only ever passes through HERA's sidelobes. It also explains the narrowing of the dual peak structure at high frequencies visible in Figure~\ref{fig:chisq_waterfall}. At higher frequencies, the beam narrows, so Fornax A enters the sidelobes later, spends less time in the main lobe, and exits the sidelobes earlier. We return to more quantitatively assess this explanation and the effect of non-redundancy on the temporal structure of the gains in Section~\ref{sec:temporal}.

The frequency dependence again shows a non-monotonic trend. Its origin is not obvious. A similar hump between $\sim$165\,MHz and $\sim$185\,MHz in $\chi^2 / \text{DoF}$ was seen by \citet{ali_et_al2015} using PAPER (see their Figure~5). Given that PAPER and HERA Phase I share feeds and signal chains, perhaps this is attributable to some property of the spectral or spatial response of the analog system. However, no comparable peaks at low or middle frequencies are seen in \citet{ali_et_al2015}. Comparing this spectrum to average absolute calibration bandpass determined for HERA in \citet{kern_et_al2019c} also shows spectral structure at roughly the same scale, but it does not appear to be correlated with the structure seen here---some dips in the bandpass are also dips in $\chi^2 / \text{DoF}$, but other dips in the bandpass are bumps in $\chi^2 / \text{DoF}$. The highest peak in $\chi^2 / \text{DoF}$ is perilously close to 137\,MHz, the frequency of the ORBCOMM constellation of satellites (seen as a spike in Figure~\ref{fig:example_cspa_outliers} which is made prior to RFI flagging and as a hole in the lower panel of Figure~\ref{fig:chisq_medians}). Given both our aggressive RFI flagging and the fact that the median should be relatively immune to occasional unflagged RFI, this explanation also seems incomplete.

Our best hypothesis is that the spectral structure in  $\chi^2 / \text{DoF}$ reflects beam directivity; the peaks and troughs are roughly aligned with those in HERA's total gain (see Figure~18 of \citet{deboer_et_al2017}). Equivalently, $\chi^2$ appears anticorrelated with beam total area $\Omega_p$ as defined in Appendix~B of \citet{parsons_et_al2013}. Naively, one might expect that higher beam directivity should lower  $\chi^2 / \text{DoF}$---relatively less sensitivity to the sidelobes should dampen the effect described above to explain the temporal structure. However, Figure~\ref{fig:chisq_waterfall} shows spectral structure that is fairly consistent in time and thus independent of the sky configurations. Perhaps greater directivity comes at the cost of more well-defined sidelobes and deeper beam nulls, which might in turn exacerbate the relative variation from antenna to antenna. This question is difficult to answer without a high-fidelity beam and source model in order to understand how much flux is in the sidelobes for a given visibility for a given time, frequency, and baseline. It merits further study beyond the scope of this work.

\subsection{Assessing the Antenna- and Baseline-Dependent Structure of $\chi^2$} \label{sec:chisq_structure}

But first, we would like to push the exploration of $\chi^2$ beyond where previous applications of redundant-baseline calibration have gone, namely to examine the breakdown of how different baseline groups and different antennas contribute to it. This discussion builds upon the mathematical framework for calculating the expectation value of $\chi^2$ for specific baselines developed in Section~\ref{sec:per-bl-dof} and on the antenna cuts based largely on $\chi^2$ that we presented in Section~\ref{sec:malfunctioning}. 

\begin{figure}
    \centering
    \includegraphics[width=.5\textwidth]{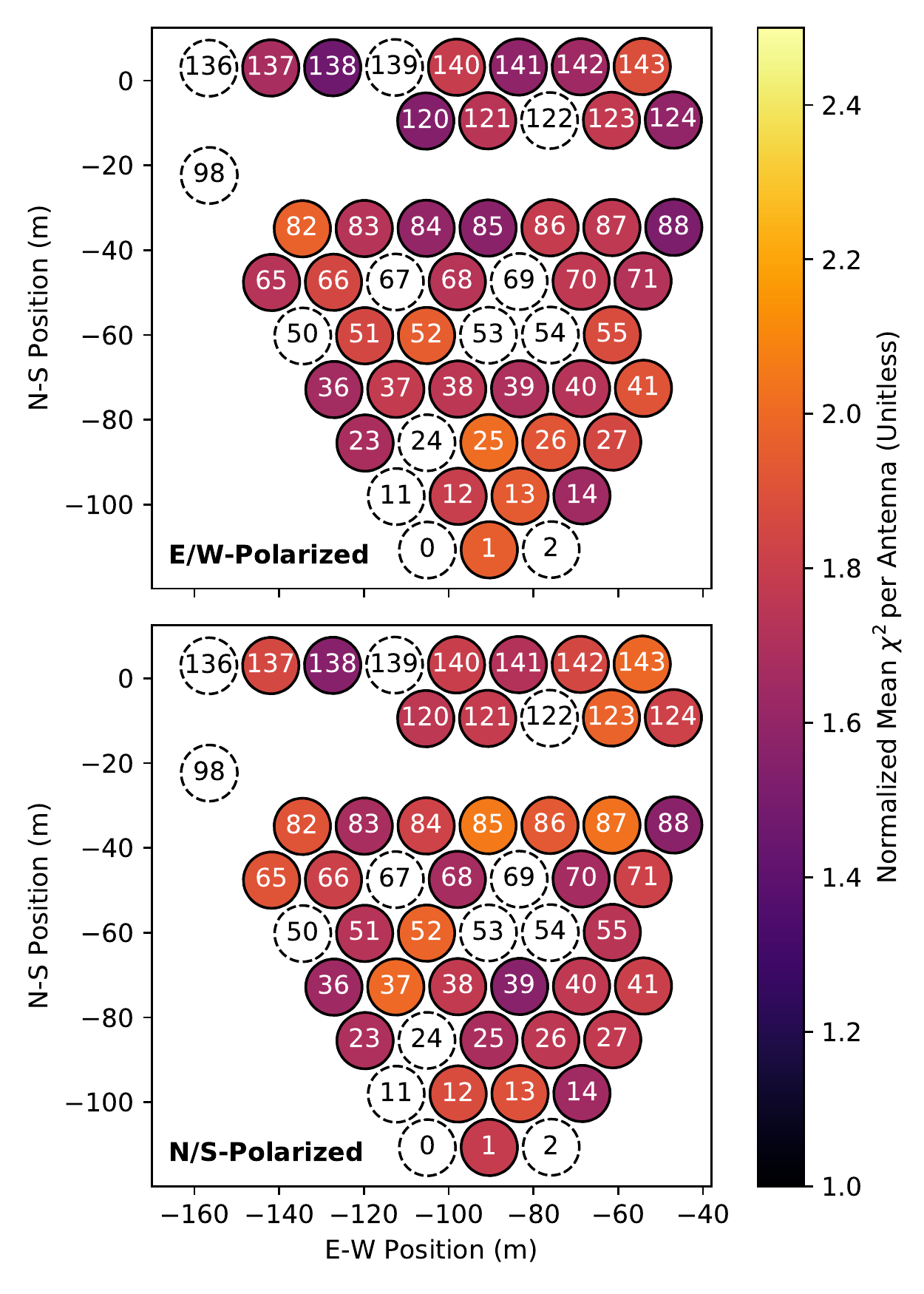} 
    \vspace{-10pt}
    \caption{Here we show our per-antenna $\chi^2$, normalized by the expected number of degrees of freedom and averaged over all unflagged times, frequencies, and nights. After having removed the strongest outliers (white dashed circles; see Section~\ref{sec:malfunctioning}), we see little clear pattern in the remaining antennas. $\chi^2$ does not appear to depend strongly on position in the array nor does it appear strongly correlated on the same antenna for the two polarizations. While a useful data quality check, this gives us limited physical understanding of the origin of our non-redundancy since the dishes have been retrofitted with new feeds since the data were taken.}
    \label{fig:obs_cspa}
\end{figure}

We start with each antenna's individual $\chi^2$, which is the sum of all the terms in the overall $\chi^2$ in Equation~\ref{eq:chisq} that involve the particular antenna, normalized by the degrees of freedom calculated using Equation~\ref{eq:chisq_per_bl}. We plot the mean values over all unflagged times, frequencies, and nights for all unflagged antennas in Figure~\ref{fig:obs_cspa}.

After removing the worst antenna outliers, little pattern remains in the antennas. One might expect, if antenna beam deformation due to neighboring antennas were a dominant source of non-redundancy, that antennas near the edge of the array would be particularly non-redundant. We see no clear evidence for this. Interestingly, we also see no significant correlation between $\chi^2$ seen by the two polarizations of each antenna, though this was not the case for the most egregious outliers, as was clearly demonstrated in Figure~\ref{fig:example_cspa_outliers} with Antennas 98 and 136. For the highest-quality antennas, this indicates that the dominant source or sources of non-redundancy are not ones that should affect the two polarizations roughly equally---like feed height or dish position errors. Rather, errors in feed horizontal positioning or orientation seem likelier culprits, since they can more easily affect the way the two polarizations differentially illuminate an imperfect dish surface. Unfortunately, since these dishes have already been retrofitted with HERA Phase II broadband feeds, there is no way to verify this hypothesis.

\begin{figure*}
    \centering
    \includegraphics[width=\textwidth]{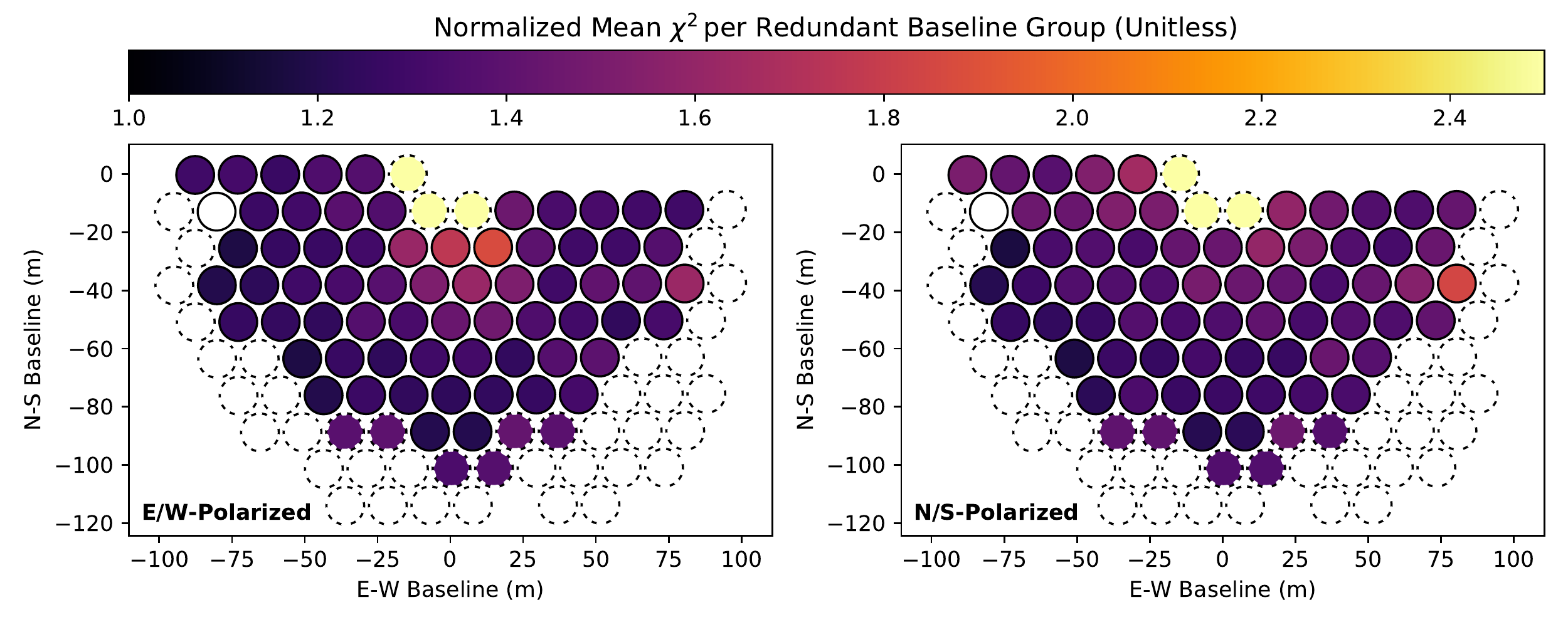} 
    \vspace{-10pt}
    \caption{Here were show $\chi^2$ for each redundant-baseline group, averaged over all unflagged times, frequencies, and nights. In our fiducial calibration scheme, baselines longer than 90\,m or shorter than 15\,m are excluded from calibration (dashed outlines) and are instead solved for afterwards. Baselines in white are unique separations, meaning that $\langle \chi^2 \rangle = 0$ and thus undefined after normalization. In general, shorter baselines have more non-redundancy structure. The precise cause of this effect is unclear, but several factors may contribute. First these baselines have larger contributions from bright but spatially smooth galactic emission, especially in the sidelobes. Relatedly, they have higher SNR, making any non-redundancy appear larger relative to the noise in $\chi^2$. They also exhibit the strongest cross-coupling systematics \citep{kern_et_al2019b}, which also source non-redundancy.}
    \label{fig:obs_cspr}
\end{figure*}

Next we examine the structure of $\chi^2$, broken down by the redundant-baseline groups involved. In Figure~\ref{fig:obs_cspr}, we show this in the case of our fiducial calibration scheme---the one used in the rest of this work.
After removing our worst antennas, we also restrict the baselines used in calibration. We exclude all baselines longer than 90\,m, motivated by the detrimental impact of long baselines on calibration spectral structure \citep{Orosz_et_al2019}. We also exclude the three shortest baselines, which show the strongest impact of cross-coupling \citep{kern_et_al2019b}. For the baselines we exclude from calibration, we apply the gain calibration solutions derived from the other baselines and then average within the redundant group to estimate the visibility solution. To normalize the result, we divide by the number of degrees of freedom that that baseline would have had if no baselines were excluded from calibration.

Figure~\ref{fig:obs_cspr} shows that the highest $\chi^2$ appears on the shortest baselines. This can be attributed, at least in part, to a number of factors. The shortest baselines have the greatest contribution from bright, but mostly spatially-smooth Galactic synchrotron emission---much of which appears in the sidelobes and thus likely varies more from antenna-to-antenna than emission in the main lobes of the primary beam (see the second panel of Figure~\ref{fig:chisq_medians}). The shortest baselines also exhibit the largest effect of temporally-stable cross-coupling systematics \citep{kern_et_al2019b}, the impact of which is unlikely to affect redundant baselines equally. The additional contribution from bright galactic emission also increases the SNR on these baselines. This increases the impact of non-redundancy $\chi^2$---by increasing the amplitude of both terms in the numerator of Equation~\ref{eq:chisq} relative to the denominator. Thus the same fractional non-redundancy produces a larger $\chi^2$ than it would on other baselines. We will return to the question of fractional non-redundancy and try to assess it in a way that is ideally independent of both noise and signal strength in Section~\ref{sec:fractional_nonred}. 

\begin{figure*}
    \centering
    \includegraphics[width=\textwidth]{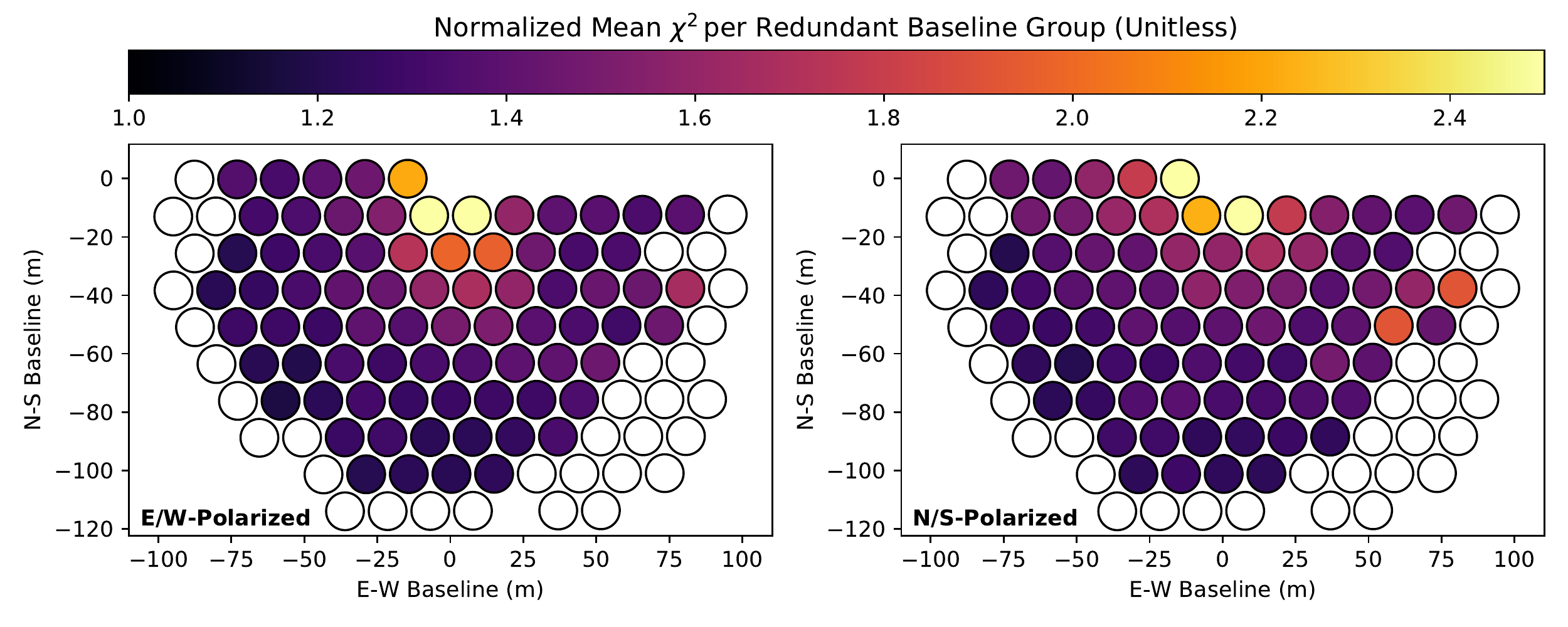} 
    \vspace{-10pt}
    \caption{Here we show the same $\chi^2$ per baseline metric we calculated in Figure~\ref{fig:obs_cspr}, this time without excluding any baselines from calibration. While $\chi^2$ on the shortest baselines is reduced relative to Figure~\ref{fig:obs_cspr}, this comes at the cost of higher $\chi^2$ across other baselines, especially those of moderate length. We can thus justify our fiducial choice of baseline cuts by arguing that is useful to keep the non-redundancy contained and not let these short baselines disproportionately affect our calibration.}
    \label{fig:obs_cspr_no_bl_cut}
\end{figure*}

That said, to some extent high $\chi^2$ on short baselines was a self-fulfilling prophecy. By excluding them from the calibration, we removed their impact on the overall minimization of $\chi^2$, effectively trading higher $\chi^2$ on short baselines for lower $\chi^2$ on all other baselines. To check this effect, we also performed redundant-baseline calibration without excluding any baselines. We show the results in Figure~\ref{fig:obs_cspr_no_bl_cut}.
Letting the shortest baseline affect the calibration raises $\chi^2$ elsewhere, especially on the moderate-length baselines. This fact, along with the reasons enumerated above, underlies our decision to exclude these baselines from our main analysis. However, it would be useful to return to this question in the future in order to tease apart the origin of non-redundancy on the short baselines to see if it can be mitigated---especially if other systematics are reduced sufficiently so that the shortest baselines can be confidently included in future power spectrum measurements \citep{kern_et_al2019b}.

\subsection{Fractional Non-Redundancy} \label{sec:fractional_nonred}

It is clear that $\chi^2$---whether overall, per-antenna, or per-baseline---is a metric that can show evidence for non-redundancy. It does not, however, answer a seemingly much simpler question: how non-redundant is HERA? Answering this question would be useful both for quantifying the deviations from ideality in the construction of our array and for comparing HERA to future redundant arrays---which is challenging to do with $\chi^2$ alone, as we saw in Section~\ref{sec:overall_dof}.

To put it another way, $\chi^2$ compares differences between the data and our redundant model of the data to the noise, but how do those differences compare to the data itself? Two factors complicate this simple question. The first is that some of that non-redundancy is due to noise and is thus ``uninteresting''---we want to know how much our visibility solutions deviate from the calibrated data beyond the variance expected from thermal noise. The second arises when we want to start comparing different baselines, times, frequencies, or nights by averaging over the other dimensions. While the noise used to normalize $\chi^2$ has relatively smooth temporal and spectral structure, that is not necessarily the case with our visibilities. When averaging together relative error measurements, we might worry that when the particular visibilities pass through destructive interference nulls, the relative error metric blows up.

We address the first problem by defining a relative error metric that looks for evidence of non-redundancy beyond that expected from noise alone. We define our estimate of the relative error on a baseline group $\eta_{i-j}$ as
\begin{equation}
    \eta_{i-j} \equiv \frac{\left|\sigma^2_{i-j,V} - \sigma^2_{i-j,N} \right|^{1/2}}{\left|V_{i-j}\right|} \label{eq:relative_error}
\end{equation}
where $\sigma^2_{i-j,V}$ is our estimate of the visibility variance in a redundant-baseline group corresponding to antenna separation $i-j$ and $\sigma^2_{i-j,N}$ is our proxy for the noise variance in that group. Since gains and visibility solutions are estimated from the same data that we now compare to those visibility solutions, our estimate of the variance should take this into account. And since that data is weighted in \texttt{omnical} by the inverse noise variance, $\sigma^{-2}_{ij}$, we use the weighted sample variance estimator:
\begin{equation}
    \sigma^2_{i-j,V} = \frac{\sum_{ij} \sigma_{ij}^{-2} \left(\frac{V_{ij}}{g_i g_j^*} - V_{i-j}\right)^2}{\sum_{ij}\sigma_{ij}^{-2} - \left(\sum_{ij}\sigma_{ij}^{-4}\middle) \right/ \left( \sum_{ij}\sigma_{ij}^{-2}\right)}. \label{eq:vis_sample_variance}
\end{equation}
Likewise, since some of the thermal noise is absorbed in the visibility solution, we use the same formula for $\sigma^2_{i-j,N}$, substituting the noise variance for the calibrated visibility difference and yielding,
\begin{equation}
    \sigma^2_{i-j,N} = \left[\sum_{ij}\sigma_{ij}^{-2} - \left(\sum_{ij}\sigma_{ij}^{-4}\middle) \right/ \left( \sum_{ij}\sigma_{ij}^{-2}\right)\right]^{-1}.
\end{equation}
Ideally, this estimator addresses our first problem by statistically isolating the portion of the observed non-redundancy due to thermal noise.

\begin{figure*}
    \centering
    \includegraphics[width=\textwidth]{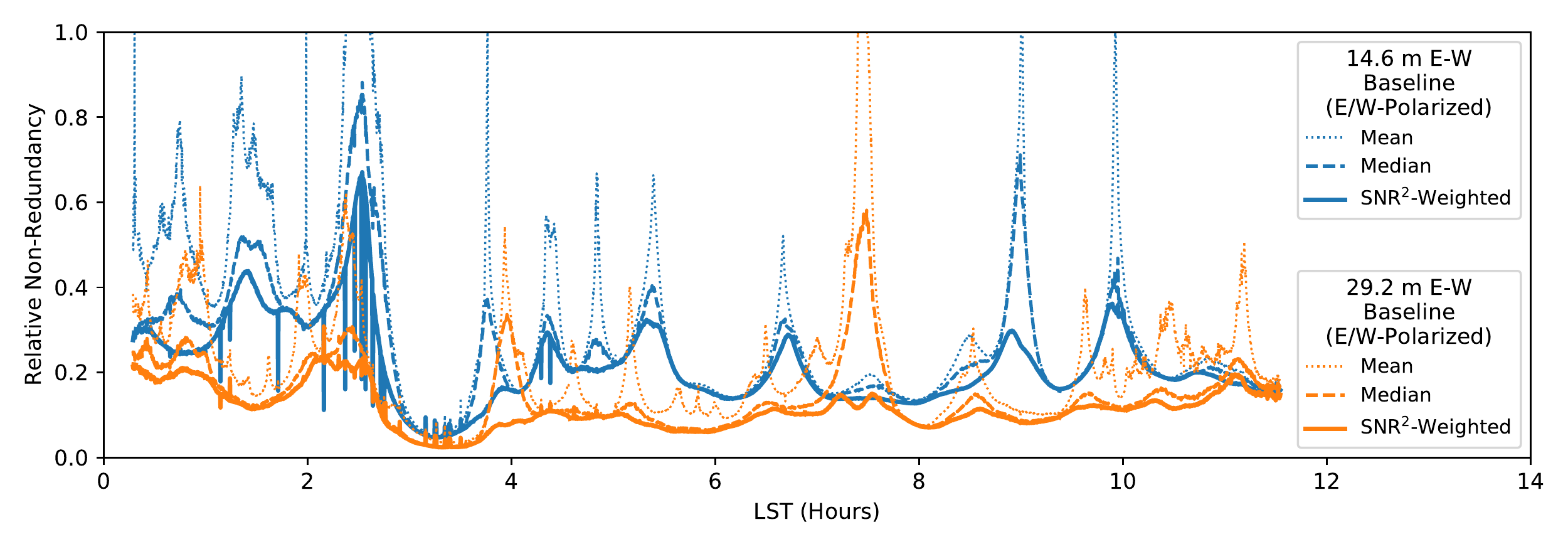} \\ \vspace{-10pt}
    \includegraphics[width=\textwidth]{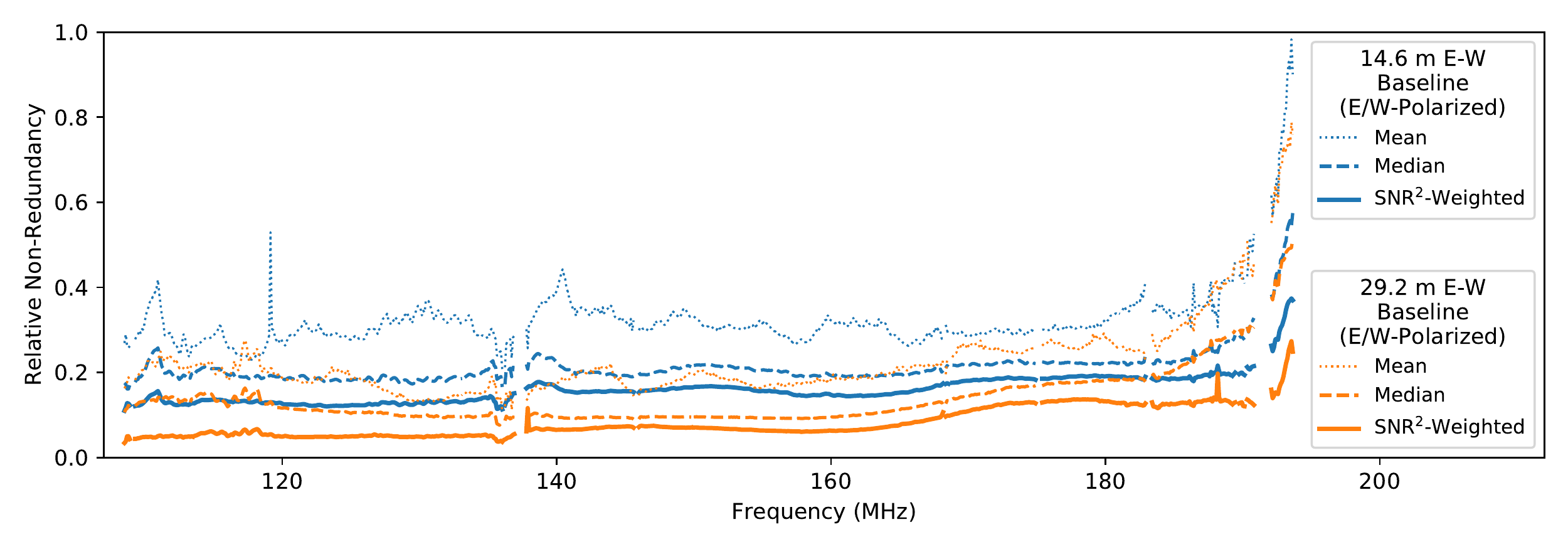} 
    \vspace{-10pt}
    \caption{Here we show our metric of relative non-redundancy in baseline groups, $\eta_{i-j}$ defined in Equation~\ref{eq:relative_error}, which was devised to separate out ``true'' non-redundancy in baseline groups from non-redundancy attributable to thermal noise. We compute $\eta_{i-j}$ for all frequencies, LSTs, nights, and baseline groups. However, reducing this statistic along one or more axes is complicated by the nulls in the visibility. Means and even medians show substantial temporal structure (top panel), largely due to this effect. If we assign the highest weight to the best-measured visibilities while averaging using Equation~\ref{eq:SNR2-weight}, this effect is smoothed out substantially. While the apparent deviations from redundancy can be as big as the visibilities themselves, in general we see $\lesssim$20\% relative non-redundancy across most of the band for short baselines and $\lesssim$10\% on longer baselines.}
    \label{fig:obs_non_red_by_lst_and_freq}
\end{figure*}

\begin{figure*}
    \centering
    \includegraphics[width=\textwidth]{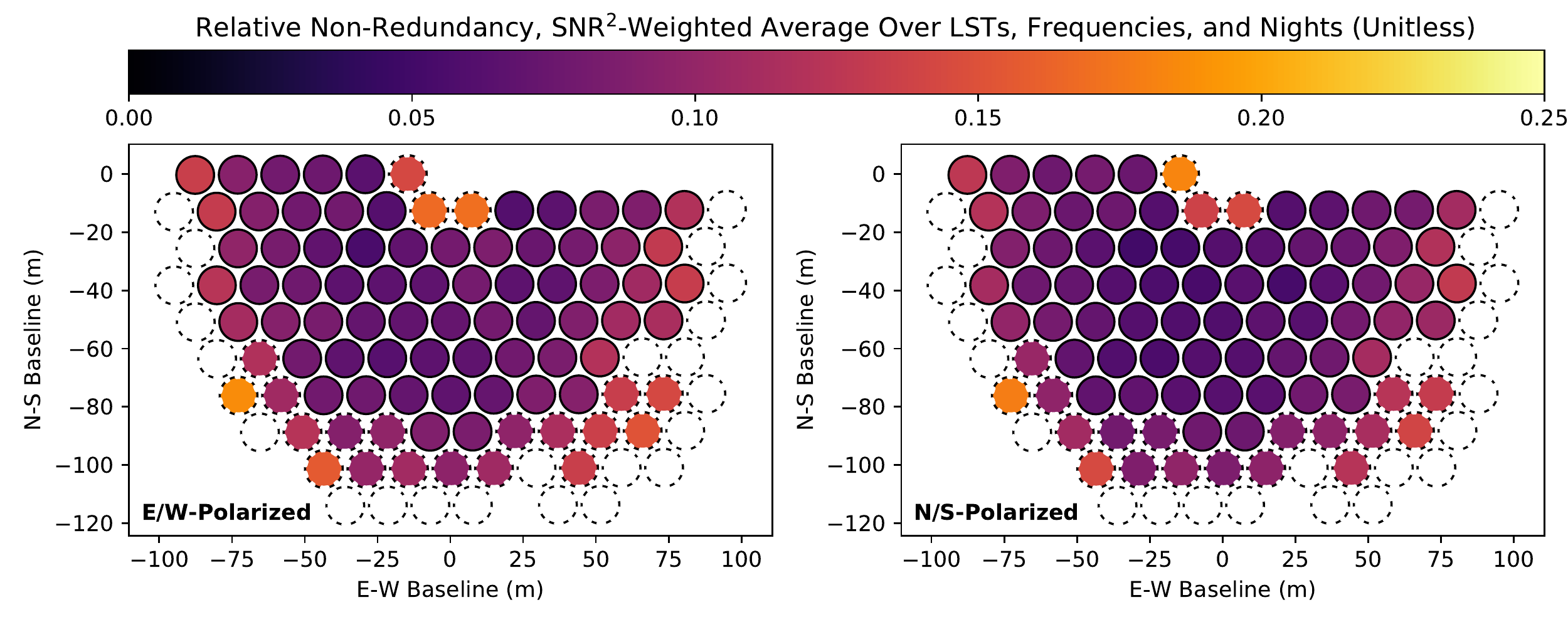}
    \vspace{-10pt}
    \caption{Our metric of relative non-redundancy of baseline groups, $\eta_{i-j}$, defined in Equation~\ref{eq:relative_error} and averaged over LST, frequency, and night using SNR$^2$-weighting (Equation~\ref{eq:SNR2-weight}). Most of our baselines show sub-10\% relative non-redundancy, consistent across both instrumental polarizations. Larger relative non-redundancy is observed for the shortest baselines and some of the longest baselines which were excluded from calibration (dashed outlines, see Section~\ref{sec:chisq_structure}). The longer baselines are also less sampled, so their estimates of visibility sample variance are themselves noisier.}
    \label{fig:obs_non_red}
\end{figure*}

The estimator of relative error $\eta_{i-j}$ in Equation~\ref{eq:relative_error} is straightforward to compute as a function of baseline, time, frequency, and night. However, when we try to form any summary statistics we immediately run into a problem. In Figure~\ref{fig:obs_non_red_by_lst_and_freq}, we plot $\eta_{i-j}$ as a function of both LST and frequency for a pair of baselines, averaging over night and either frequency or LST.
When the amplitude of the visibility is low, this statistic swings wildly, sometimes yielding relative error estimates of 10 or more. These times and frequencies end up dominating the average and sometimes even the median. This heavy variability in time, seen most clearly in the top panel of Figure~\ref{fig:obs_non_red_by_lst_and_freq}, produces mean estimates of $\eta$ well above the medians. 
Since our goal is to develop a somewhat array-independent metric of redundancy, it seems odd for the quantity to depend so strongly on the particular time of observation. The array is not changing substantially during this time, so do we really trust all of these estimates of $\eta_{i-j}$ equally? We propose a weighting scheme for averaging $\eta_{i-j}$ where each estimate is weighted by 
\begin{equation}
    w_{i-j} \equiv \frac{\left|V_{i-j}\right|^2}{\sigma^2_{i-j,N}}. \label{eq:SNR2-weight}
\end{equation}
This SNR$^2$-weighting gives the most weight to the visibilities measured with the highest signal to noise and removes the undue influence of visibility nulls. In Figure~\ref{fig:obs_non_red_by_lst_and_freq} our SNR$^2$-weighting produces substantially smoother estimates of $\langle \eta_{i-j} \rangle$ for both baselines plotted. We use this weighting to investigate how non-redundancy depends on baseline group in Figure~\ref{fig:obs_non_red}.

With the exception of the shortest and longest baselines, most baseline groups show non-redundancy at the sub-10\% level. This result is consistent with other metrics of non-redundancy, e.g.\ closure phase \citep{Carilli_et_al2018}. Likewise, this is consistent with the rough level of non-redundancy seen in the fiducial simulations of \citet{Orosz_et_al2019}, which take as inputs to their parameterized non-redundancy the construction tolerances of HERA dish and feed positioning. The longest baselines are the most infrequently measured, so the estimate of the sample variance in Equation~\ref{eq:vis_sample_variance} is likely quite noisy. Similarly, we expect the shortest baselines to be the least-redundant, in part due to their sensitivity to diffuse galactic structure in the sidelobes, which likely vary more from antenna-to-antenna than other parts of the beam.  Still, they are only non-redundant at the $\sim$20\% level, which is encouraging given that they were excluded from the determination of the gains in redundant-baseline calibration.

%%%%%%%%%%%%%%%%%%%%%%%%%%%%%%%%%%%%%%%%%%%%%%%%%%%%%%%%%%%%%%%%%%%%%%%%%%%%%%%%%%%%%%%%%%%%%%%%%%%%%%%%%

\section{The Effect of Non-Redundancy on Calibration Temporal Structure} \label{sec:temporal}

Because spectral smoothness is so key to 21\,cm cosmology, one may worry that any calibration using incomplete knowledge---be it sky or beam knowledge in sky-based calibration \citep{barry_et_al2016, ewall-wice_et_al2017} or unknown deviations from redundancy \citep{Orosz_et_al2019, Byrne_et_al2019}---might impart spurious spectral structure on calibration solutions. We see clear evidence for non-redundancy at a level comparable to the fiducial $\sim$10\% error level in \citet{Orosz_et_al2019}, so this concern appears to be pressing. However, it does not appear to be the leading-order contribution to unsmooth gains; \citet{kern_et_al2019c} shows clear evidence for spurious spectral structure in HERA's solved gains using both sky-based and redundant-baseline calibration attributable to cross-coupling systematics \citep{kern_et_al2019a, kern_et_al2019b}. As was suggested in the conclusion of \citet{Orosz_et_al2019}, \citet{kern_et_al2019c} demonstrate that low-pass filtering of calibration solutions appears to be a robust way of mitigating this effect. This approach follows a ``do-no-harm'' philosophy, relying on the instrument to impart less spectral structure than we would with analysis errors. Whether that approach can carry us through to a detection and characterization of the 21\,cm cosmological signal remains to be seen. 

While \citet{kern_et_al2019c} has thoroughly explored the spectral structure of calibration solutions, they have largely set aside the effects of non-redundancy on temporal structure. As we will show in Section~\ref{sec:obs_temp}, we see clear evidence for temporal structure well in excess of the expected gain drifts with ambient temperature. In this final section of the paper, we quantify the apparent temporal structure in our calibration solutions and attempt to assess how much of it is real and how much of it is an artifact of non-redundancy, invoking analogous simulations to those of \citet{Orosz_et_al2019} in Section~\ref{sec:explain_temp}. This study lets us motivate future temporal filtering of gains, analogous to how \citet{zheng_et_al2014} used the Fourier-space statistics of the calibration solutions to develop a Wiener filter kernel.

\subsection{Observed Temporal Structure in Gains} \label{sec:obs_temp}

\begin{figure*}
    \centering
    \includegraphics[width=\textwidth]{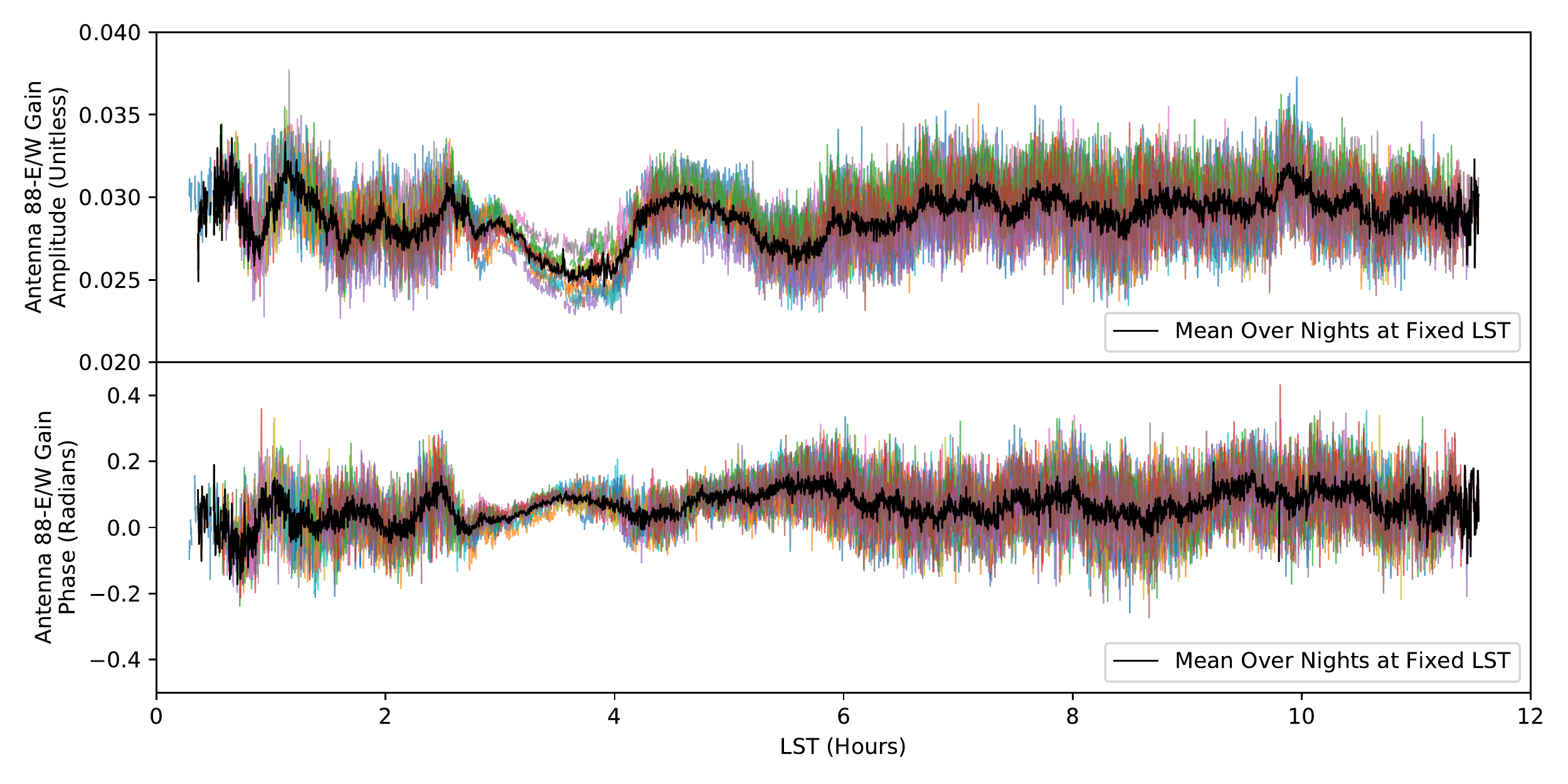}
    \vspace{-10pt}
    \caption{One consequence of non-redundancy in redundant-baseline calibration is the introduction of calibration errors which depend strongly on the configuration of sources in the beam at a given LST and thus repeat from night to night. Plotting our calibration solutions as a function of LST for a single antenna and frequency for all nights demonstrates this effect. Here we show gain amplitudes and phases for a single antenna---in this case, the East/West polarization of antenna 88---at a single frequency ($\sim$125\,MHz). Each different colored line is a different night in our data set after redundant-baseline calibration, absolute calibration to fix the degeneracies, and RFI flagging. The $\sim$20\% level fluctuations is typical of antennas and frequencies. While some of that fluctuation is expected from thermal noise, but the coherent fluctuations that repeat from day to day (especially around the transit of Fornax A at $\sim$4 hours) at the same LST indicate systematic error. In black we show the average across night after excluding daytime data and the most RFI-contaminated nights (2458104, 2458105, and 2458109). To avoid temporal structure in the degenerate subspace, we apply absolute calibration after redundant-baseline calibration. Since absolute calibration was performed only a few fields and then transferred to whole nights \citep{kern_et_al2019c}, we do not expect it to contribute significantly to the temporal structure seen here.}
    \label{fig:gains_vs_lst}
\end{figure*}

Temporal structure in calibration solutions, in and of itself, is not evidence for non-redundancy. Signal chain elements can be sensitive to temperature, for example. The real smoking gun that something is amiss is that the variability in our gains repeats from night to night at fixed LST---just as $\chi^2$ did (see Figure~\ref{fig:chisq_medians}). In Figure~\ref{fig:gains_vs_lst} we show calibration gains after both redundant-baseline calibration and absolute calibration. 
The procedure for absolute calibration is detailed in \citet{kern_et_al2019c}. We include absolute calibration to eliminate temporal discontinuities due to antenna flagging. However, because absolute calibration was performed on a single field and transferred to an entire night, we expect most of the temporal structure to be attributable to calibration errors introduced by redundant-baseline calibration.

While the calibration solutions shown in Figure~\ref{fig:gains_vs_lst} vary from day to day due to noise, there is clear evidence for repeated structure. Perhaps unsurprisingly, the strongest apparent variations occur during the transit of Fornax A, further evidence for our hypothesis that bright point sources, combined with antenna-to-antenna beam variation cause redundant gains to be erroneously dragged around. The phases are contiguous, so this does not appear to be an LST-dependence due to phase wrapping as was simulated in \citet{Joseph_et_al2018}. The particulars of the peaks and troughs vary from antenna to antenna and frequency to frequency, but the general features seen in the Figure~\ref{fig:gains_vs_lst} are representative. This raises a question: with the exception of the temporal variation which repeats from night to night, do we see any evidence for real temporal structure not attributable to noise or non-redundancy?

To study this problem statistically and understand the relevant timescales in the problem, we produce the temporal power spectra of all of our gains at a particular frequency---in our case $\sim$125\,MHz, though the results are similar across the band. To interpolate over RFI gaps, we use the iterative deconvolution algorithm of \citet{parsons_backer2009}, smoothing the calibration solutions on a 1\,minute timescale. We do the same for the nightly average (black line in Figure~\ref{fig:gains_vs_lst}). Assuming that our gain solutions are separable as the product of a LST-dependent systematic and time-dependent intrinsic fluctuations, we divide out the nightly averages, producing our best estimate of ``intrinsic'' gain variability. These continuous gains and gain ratios are then tapered with a Hann window, Fourier transformed, squared, and then re-normalized to peak at unity. 

\begin{figure*}
    \centering
    \includegraphics[width=\textwidth]{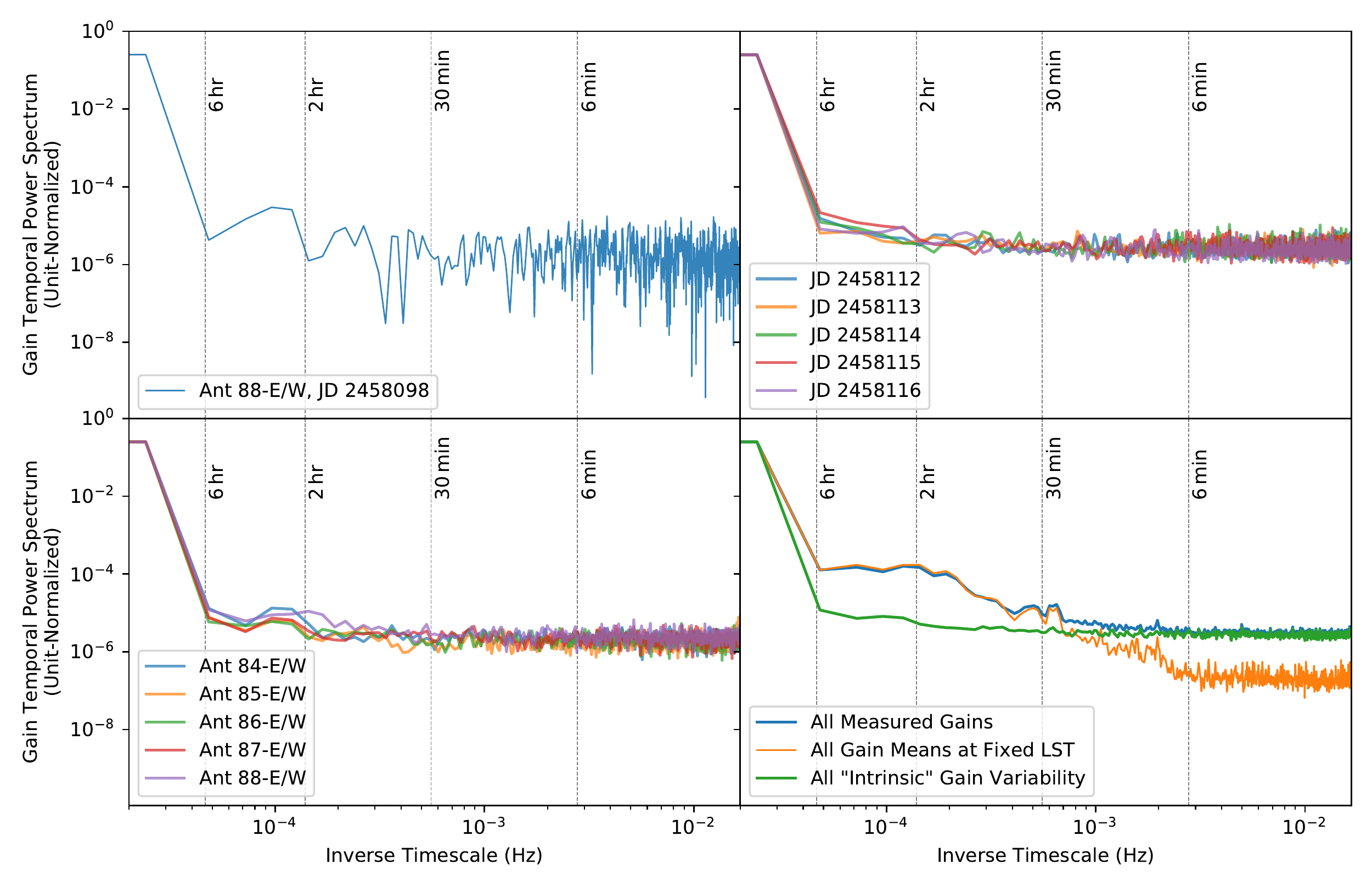}
    \vspace{-10pt}
    \caption{We compute temporal power spectra of our gains to look for ``intrinsic'' variability beyond the LST-locked component we saw in Figure~\ref{fig:gains_vs_lst} and attributed to non-redundancy. First we show the power spectrum of the same antenna as in Figure~\ref{fig:gains_vs_lst} after dividing out the average over all nights from a gains on a single night, 2458098 (top left panel). To look for overall trends, we average these power spectra incoherently over E/W-polarized antennas, plotting a selection of the results as a function of night (top right), and over nights, plotting a selection of the results as a function of antenna (bottom left). While these averages produce a less noisy estimate of the temporal power spectra, they reveal little structure beyond 6\,hours. Because the average is incoherent, the noise floor remains fixed. Finally, in the bottom right panel, we compare temporal power spectra before (blue) and after (green) dividing out the nightly average gain, in both cases averaging gain power spectra incoherently over nights and antennas. The discrepancy between the two is the spectral structure attributed to systematics that repeat from night to night at fixed LST, most likely non-redundancy. As a cross-check, we also plot the temporal power spectrum of the nightly averages, averaged incoherently over antennas (orange). As expected, this matches the observed temporal structure of the measured gains at long and medium timescales, but drops to a lower noise floor because the nightly average is coherent. The lack of structure at timescales shorter than 6\,hours justifies smoothing on that timescale, since it will not eliminate any real temporal structure in the gains above the $\sim$10$^{-3}$ level.}
\label{fig:temporal_ps} 
\end{figure*}

We show our temporal power spectrum results in stages in Figure~\ref{fig:temporal_ps}.
In general, we find that dividing out the nightly average removes all apparent spectral structure on timescales shorter than 6\,hours, at least up to a noise floor at the $\sim$0.1\% level in the gains. This appears to be true over antennas and nights, and is confirmed by the fact that the temporal power spectrum of the nightly averages themselves contains the same temporal structure as in the original redundant-calibration gains. Given the short period of observation (18 days), it is possible that some of the apparently LST-locked behavior is attributable to cyclical changes in local conditions, such as the effect of ambient temperature on gains or HERA dish/feed structure. We expect this effect to be more pronounced in the antenna-independent overall amplitude of the gains---a degeneracy of redundant-baseline calibration---but we cannot completely rule out local effects on timescales between 6 hours and the $\sim$1\,hour that LST changes relative to local time over the observing campaign.

This result leads us to the conclusion that we can safely smooth our calibration solutions on 6-hour timescales without losing any real temporal structure above that $\sim$0.1\% level. To be clear, this does not mean that our calibration solutions are correct to $\sim$0.1\%---only that any errors above that level are isolated to timescales longer than 6\,hours. 

\subsection{Explaining Temporal Structure in Terms of Non-Redundancy} \label{sec:explain_temp}

Without precise knowledge of antenna-to-antenna beam variation, it is not possible to predict the precise way in which redundant-baseline calibration of a not-quite redundant array will cause antenna gains to vary in time. Those measurements are quite challenging and are the focus of other work (e.g.\ \citealt{Jacobs_et_al2017, Nunhokee_et_al2020}). However, given the statistical understanding we have developed in Section~\ref{sec:obs_temp} of the imprint of non-redundancy on redundant-baseline calibration's gains, it is reasonable to ask whether that structure is reproducible in a relatively simple simulation. 

To test this idea, we simulate visibilities at 125\,MHz using the same set of antennas and the same range of LSTs as those observed. Our sky model consists of four components. First we use the Global Sky Model of diffuse emission \citep{deoliveira2008} at a HEALPix resolution of \texttt{NSIDE}$=128$. We then add the 1000 brightest beam-weighted point sources from the GLEAM catalog \citep{hurley_walker_et_al2017} and all of the bright radio sources that were peeled from that catalog but included in Table~2 of \citet{hurley_walker_et_al2017}. Lastly, since Fornax A was excluded from the GLEAM catalog but not included in Table~2, we model it as a single 750\,Jy point source at 150\,MHz and extrapolate to a 125\,MHz using a spectra index $\alpha=-.81$ \citep{McKinley_et_al2015}. Since we are only interested in a statistical comparison of the effect on redundant-baseline calibration, accurate sky modeling is not strictly necessary. Source mismodeling and double-counting are unlikely to affect our temporal power spectra substantially.

To simulate non-redundancy, we adopt the simplified model of HERA developed in \citet{Orosz_et_al2019}. In our simulation, antennas are perturbed from their ideal positions by Gaussian random displacements with $\sigma = .03$\,m in both directions. Our beams are Airy functions parameterized by pointing errors ($\sigma = 0.15^\circ$) and beam FWHM errors ($\sigma = 0.28^\circ$) in both directions. There is no statistical difference between the two polarizations; the ``average'' beam is circular. For mathematical details, see Section~2.1 of \citet{Orosz_et_al2019}. We simulate visibilities with perfect calibration, but then allow \texttt{omnical} to move us away from gains of 1.0 in its attempt to minimize $\chi^2$. 

\begin{figure*}
    \centering
    \includegraphics[width=\textwidth]{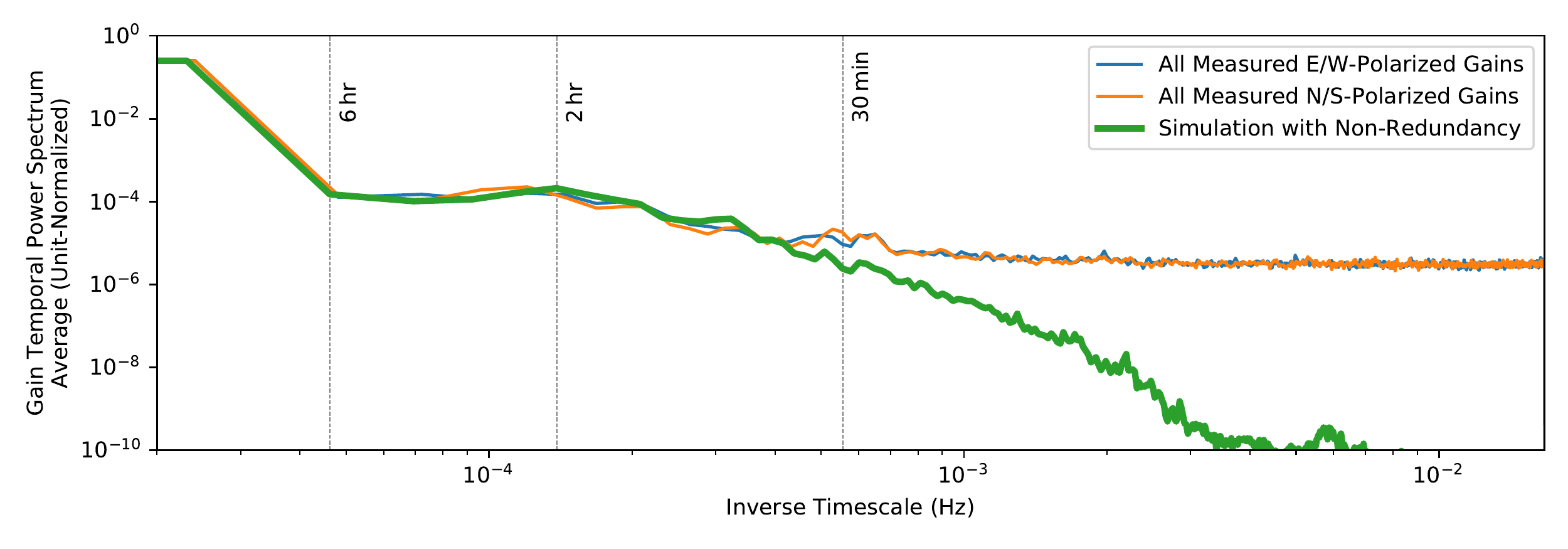}
    \vspace{-10pt}
    \caption{Comparing our simulation of the effect of non-redundancy on gain temporal power spectra to the statistics of our observed HERA gains reveals good agreement on timescales longer than a beam-crossing time. Here we show our measured gain temporal power spectra averaged incoherently over antennas and nights (the blue line is the same as the blue line in Figure~\ref{fig:temporal_ps}). At shorter timescales they disagree because our observed gains hit a thermal noise floor while the simulation is noise-free.  The agreement on long timescales demonstrates the plausibility of our hypothesis that the difference between the observed and ``intrinsic'' gain power spectra in Figure~\ref{fig:temporal_ps}---and thus all observed temporal structure on timescales shorter than 6\,hours---is attributable to the effect of non-redundancy on redundant-baseline calibration.
    }
    \label{fig:temporal_sim}
\end{figure*}

By renormalizing our gain temporal power spectra to peak at unity, we can compare our simulation directly to the observed gain variability. In Figure~\ref{fig:temporal_sim} we show how our calibration of simulated data with non-redundancy compares with our real calibration solutions. 
On timescales longer than the beam-crossing time ($\sim$40\,minutes), our simulation matches the observed temporal power spectra very well. This makes sense since over a beam-crossing time, calibration errors due to, for example, bright point sources moving through non-redundant sidelobes should be highly correlated, contributing minimally to temporal variability. Likewise, we expect our simulation to diverge from the data on short timescales because the data hits a noise floor and the simulation is noise-free. There is tentative evidence for minor disagreement at intermediate timescales---the simulated power spectrum appears to be falling a bit faster than the data. This could be attributed to the simplicity of the beam model, which probably has less spatial structure than the real beam and thus produces more correlated gain variations on timescales associated with sources passing through beam substructure, though that conclusion is rather speculative.

To be clear, our simulation is not a fit to the data. By simple guess-and-check we found remarkably good agreement with HERA data when using error levels 50\% higher than the ``fiducial'' error level in \citet{Orosz_et_al2019}. In that work, the fiducial errors produced visibility variances at the $\sim$10\% level on the shortest baselines, so increasing that by 50\% is entirely in line with the non-redundancy of HERA we observed and quantified in Section~\ref{sec:hera_redundancy}. We did not attempt to measure the levels of individual types of non-redundancy individually. Likely the effect of these different error types is highly degenerate in their effect on the gain temporal power spectra and we know that this simple 6-parameter model for each antenna does not capture the full complexity or variability of HERA elements. Despite all that, the simulation makes it clear that the observed LST-locked temporal structure in our gain solutions is largely, if not entirely, attributable to the non-redundancy we observe in HERA.

%%%%%%%%%%%%%%%%%%%%%%%%%%%%%%%%%%%%%%%%%%%%%%%%%%

\section{Summary}

In this work, we comprehensively survey the method of redundant-baseline calibration and its application to HERA. This includes a pedagogical review of the mathematical and algorithmic underpinnings of the technique, a revised quantification of the statistical expectation of $\chi^2$ (correcting a minor error in the literature), and a new formalism for predicting how different baselines, antennas, or redundant-baseline groups contribute to $\chi^2$. We apply this technique to HERA data, producing redundant-baseline calibration solutions that solve for most of the internal degrees of freedom
and enable future absolute calibration to fix the last few degenerate modes. \citet{kern_et_al2019c} show that that process works well with HERA and that the combination of redundant-baseline and subsequent absolute calibration compares favorably to pure sky-based calibration with HERA. That said, both techniques introduce spectral structure into the calibration solutions---likely attributable to cross-talk systematics---that must be filtered on delay scales larger than $\sim$100\,ns. Exploring methods of redundant-baseline calibration that are immunized to these sorts of systematics remains an active area of research.

We also use redundant-baseline calibration to assess the health of the array and study the origins of non-redundancy---an important systematic for 21\,cm cosmology that affects calibration and can adversely narrow the EoR window if not taken into account \citep{Orosz_et_al2019, Byrne_et_al2019}. We study $\chi^2$---the quantity redundant-baseline calibration seeks to minimize---and show how particular antennas contribute disproportionately to it, likely pointing to some flaw in their construction. We find the largest deviations from redundancy on short baselines, both in terms of $\chi^2$ and in terms of relative error. We attribute them to some combination of larger cross-talk effects and higher sensitivity to diffuse emission in the sidelobes the array which are believed to be relatively more non-redundant than the main lobe of the primary beam. By one metric, we find that almost all of our putatively redundant-baseline groups show visibility variance at or below the 10\% level, roughly in line with our expectations given the construction tolerances of element construction and placement. To study this, we develop a metric for non-redundancy that is ideally comparable between HERA and other redundantly arranged telescopes and largely insensitive to the specific sky configuration within the beam. 

Since \citet{kern_et_al2019c} already studied the spectral structure of redundant-calibration solutions in detail, we then turn to an assessment of the impacts of redundant-baseline calibration on the temporal structure of calibration solutions. Though we see substantial temporal structure in gain solutions, we find that it repeats from night-to-night at fixed LST, indicating a systematic effect. Taking out the nightly averages, we find little evidence for intrinsic variation temporal structure on timescales shorter than 6\,hours, justifying a long smoothing timescale. Inspired by \citet{Orosz_et_al2019}, we simulate this effect in a simplified model of the array and find that the entire effect can be explained by non-redundancy at roughly the level we see in HERA.

Overall, we find that redundant-baseline calibration is a powerful tool for making sense of data from a new array without the requirement of substantial prior knowledge about that array or even the sky at the observed frequencies. While redundant-baseline calibration is vulnerable to systematics introduced by non-redundancy, so too is sky-based calibration if that non-redundancy is not precisely measured and forward-modeled. Nevertheless, considerable progress has already been made on understanding the origin and nature of these systematics and a number of ideas---from calibrating with only relatively short baselines to filtering calibration solutions---have been proposed to mitigate them. Hopefully, the tools we have detailed, refined, and applied to HERA here will continue to serve HERA and future 21\,cm arrays in quantifying and avoiding systematics in the quest to separate bright astrophysical foregrounds from high-redshift 21\,cm signal.

%%%%%%%%%%%%%%%%%%%%%%%%%%%%%%%%%%%%%%%%%%%%%%%%%%

\section*{Acknowledgements}
Work similar to that presented here appeared previously in unrefereed formats in both public HERA Memos \#50 \citep{hera_memo_50}, \#61 \citep{hera_memo_61}, \#69 \citep{hera_memo_69}, and \#72 \citep{hera_memo_72}, as well as in Zaki Ali's University of California, Berkeley Ph.D.\ thesis \citep{Ali:thesis}. This work was enabled by a number of software packages including \texttt{scipy} \citep{2020SciPy-NMeth}, \texttt{numpy} \citep{numpy}, \texttt{matplotlib} \citep{Hunter:2007}, \texttt{astropy} \citep{astropy:2013, astropy:2018}, and  \texttt{pyuvdata} \citep{hazelton_et_al2017}.

This material is based upon work supported by the National Science Foundation under grants \#1636646 and \#1836019 and institutional support from the HERA collaboration partners. 
This research is funded in part by the Gordon and Betty Moore Foundation. 
HERA is hosted by the South African Radio Astronomy Observatory, which is a facility of the National Research Foundation, an agency of the Department of Science and Innovation. 
JSD gratefully acknowledges the support of the NSF AAPF award \#1701536 and the Berkeley Center for Cosmological Physics. JSD also wishes to acknowledge a helpful discussion with Martin White that led to a number of clarifications to the Algorithm~\ref{alg:omnical} and the surrounding prose.
A.\ Liu acknowledges support from a Natural Sciences and Engineering Research Council of Canada (NSERC) Discovery Grant and a Discovery Launch Supplement, as well as the Canadian Institute for Advanced Research (CIFAR) Azrieli Global Scholars program. 
Parts of this research were supported by the Australian Research Council Centre of Excellence for All Sky Astrophysics in 3 Dimensions (ASTRO 3D), through project number CE170100013. 
GB acknowledges funding from the INAF PRIN-SKA 2017 project 1.05.01.88.04 (FORECaST), support from the Ministero degli Affari Esteri della Cooperazione Internazionale - Direzione Generale per la Promozione del Sistema Paese Progetto di Grande Rilevanza ZA18GR02 and the National Research Foundation of South Africa (Grant Number 113121) as part of the ISARP RADIOSKY2020 Joint Research Scheme, from the Royal Society and the Newton Fund under grant NA150184 and from the National Research Foundation of South Africa (Grant No.\ 103424).
MGS acknowledges support from the South African Radio Astronomy Observatory (SARAO) and the National Research Foundation (Grant No.\ 84156).

%%%%%%%%%%%%%%%%%%%%%%%%%%%%%%%%%%%%%%%%%%%%%%%%%%

\section*{Data availability}
The data underlying this article are stored at the Array Operations Center of the National Radio Astronomy Observatory in Socorro, New Mexico. The data are quite large ($\sim$10\,TB), but can be made available to interested researchers upon reasonable request, subject to HERA collaboration policies.


\bibliographystyle{mnras}
\bibliography{biblio} % if your bibtex file is called example.bib

%%%%%%%%%%%%%%%%%%%%%%%%%%%%%%%%%%%%%%%%%%%%%%%%%%

%%%%%%%%%%%%%%%%% APPENDICES %%%%%%%%%%%%%%%%%%%%%

\appendix

\section{Methods For Estimating Delays from Gains or Visibilities} \label{sec:fft_dly}

In redundant-baseline calibration, precise determination of antenna delays---the dominant term in the phase of antenna gains---helps later iterative steps (e.g.\ \texttt{omincal}) converge faster and more reliably. It also avoids the complication of adding spectral structure in the degenerate subspace of redundant-baseline calibration that must be later removed via absolute calibration \citep{dillon_et_al2017}. Determining those delays via \texttt{firstcal} (see Section~\ref{sec:firstcal}) requires estimating delays from visibilities or visibility products, as in Equation~\ref{eq:firstcal}. Before solving a system of equations, as in \texttt{firstcal}, we must determine delays $\tau$ from complex data products $d(\nu)$ (be they gains, gain products, visibilities, visibility products, etc.) that look like
\begin{equation}
    d(\nu) \approx d_0 e^{2\pi i \nu \tau}. \label{eq:delay_eq}
\end{equation}

\begin{figure}
    \centering
    \includegraphics[width=.5\textwidth]{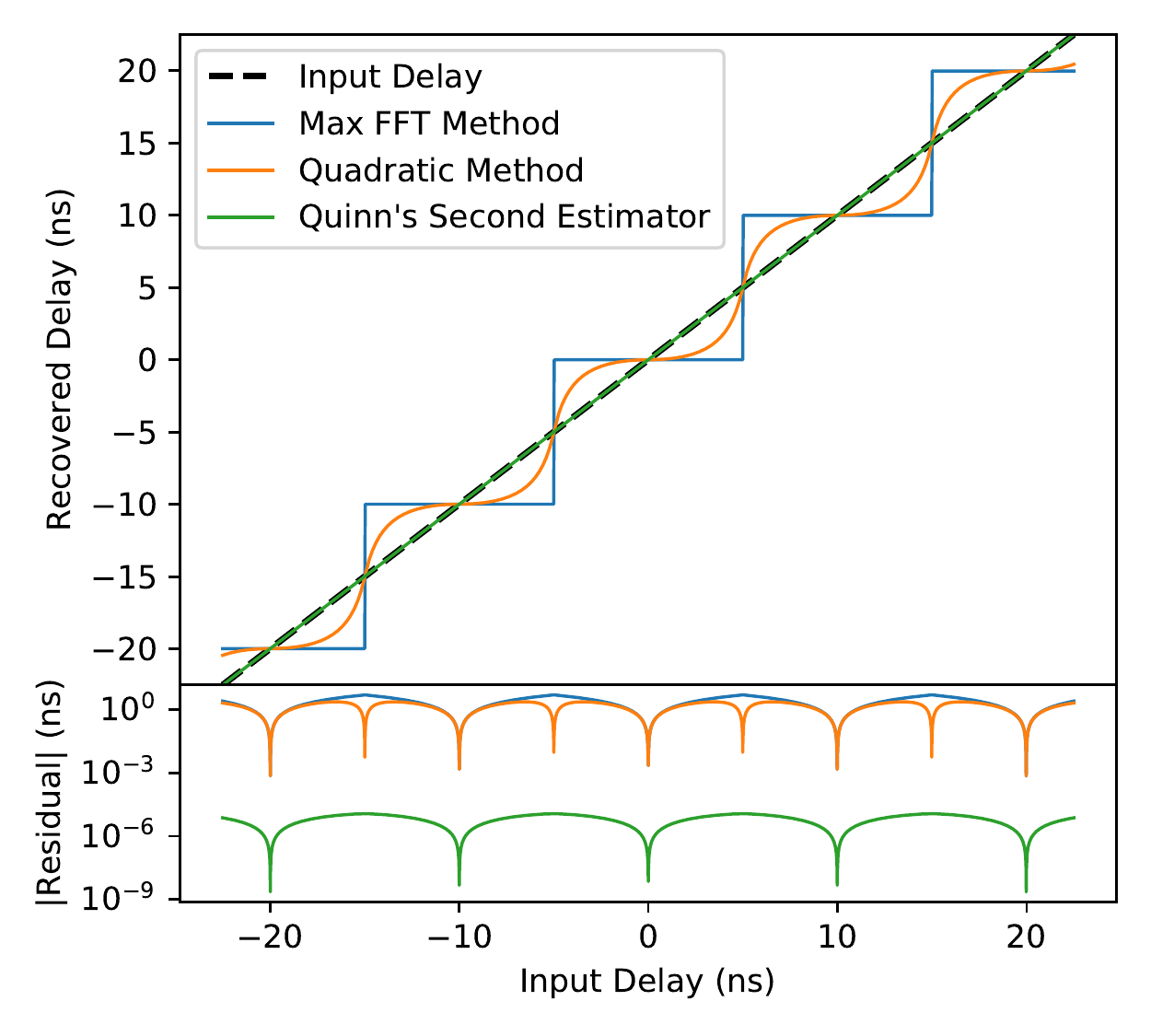}
    \vspace{-10pt}
    \caption{In general, one can estimate a delay in a spectrum by looking the delay where the absolute value of the FFT peaks. Here we compare that that method (blue lines) to two variants that use information about the two neighboring values of the FFT. In the quadratic method (orange lines), those three points are interpolated with a parabola to find the peak $\tau$. This is an improvement, but even more sophisticated methods exist in the signal processing literature. We use Quinn's Second Estimator \citep{Quinn1997}, which we explain quantitatively in Appendix~\ref{sec:fft_dly}. The method uses the same three pieces of information---the peak of the FFT and its neighbors---to produce results orders of magnitude more accurate in our noise-free demonstration (green lines). All three methods produce perfect delays when the input $\tau$ is an integer multiple of $1/B$, where $B$ is the bandwidth (in our test, 100\,MHz).}
    \label{fig:fft_dly}
\end{figure}

To the extent that Equation~\ref{eq:delay_eq} holds, the problem of estimating $\tau$ from $d(\nu)$ is equivalent to finding the peak of the Fourier transform of $d(\nu)$. The simplest peak finding algorithm is to take the FFT of $d(\nu)$, which we define as $\widetilde{d}(\tau)$, and find the delay corresponding to the maximum absolute value. In an measurement with 100\,MHz of bandwidth, this technique yields delay resolution of 10\,ns and thus errors as large as 5\,ns (see Figure~\ref{fig:fft_dly}). As \citet{dillon_et_al2017} showed, delay errors that large create phase wraps that add spectral structure in the phase degeneracies. 

This problem is well-studied in the signal processing literature, where it is usually framed as the problem of \emph{frequency estimation} from time series data. There exist other, more accurate algorithms that are not more computationally intensive than taking an FFT. Perhaps the next simplest is to find the peak absolute value of the FFT and interpolate between it and two nearest neighbors with the unique parabola that describes the three points. The interpolated peak is thus the maximum value of the parabola, which is shifted from the maximum value of the FFT toward the larger of the two neighbors. This technique still has the virtue of computational simplicity and locality in delay space. In the simple case of a single input tone, Figure~\ref{fig:fft_dly} shows that this method produces smaller errors than the maximum FFT approach.

Interestingly, it is possible do substantially better just using the same three pieces of information---the peak of the FFT and its two neighbors---if ones uses both their real and imaginary parts rather than taking the absolute value. One such method is \emph{Quinn's Second Estimator} \citep{Quinn1997}. Since this method has not, to best of our knowledge, been used in the radio astronomy literature, we reproduce it concisely here. If $\widetilde{d}(\tau)$ is maximized at $\tau_0$ for the discrete set of delays produced by the FFT, and if the two neighboring delays are denoted $\tau_{-1}$ and $\tau_{+1}$, each $\Delta \tau$ from $\tau_0$, then Quinn's Second Estimator of delay, $\widehat{\tau}$, is given by \begin{equation}
    \widehat{\tau} \equiv \tau_0 + \Delta \tau \left[ \frac{\delta_{-1} + \delta_{+1}}{2} + \kappa\left(\delta_{-1}^2 \right)  - \kappa\left(\delta_{+1}^2 \right)\right].
\end{equation}
Here the $\delta_{\pm 1}$ terms are defined as
\begin{equation}
    \delta_{\pm 1} \equiv \frac{\mp \Real{\widetilde{d}\left( \tau_{\pm 1} \right) / \widetilde{d}\left( \tau_0 \right)}}{1 - \Real{\widetilde{d}\left( \tau_{\pm 1} \right) / \widetilde{d}\left( \tau_0 \right)}}, 
\end{equation}
and $\kappa(x)$ is defined as 
\begin{equation}
    \kappa(x) \equiv \frac{1}{4} \ln \left(3 x^2 + 6x + 1) \right) - \frac{\sqrt{6}}{24} \ln \left( \frac{x + 1 - \sqrt{2/3}}{x + 1 + \sqrt{2/3}} \right).
\end{equation}
While mathematically more complicated (and certainly far less intuitive), this estimator is simple to compute and performs orders of magnitude better than the quadratic method in the simple scenario of a single delay with no noise (Figure~\ref{fig:fft_dly}). It is also, as \citet{Quinn1997} shows, a robust estimator in the presence of noise.

%%%%%%%%%%%%%%%%%%%%%%%%%%%%%%%%%%%%%%%%%%%%%%%%%%

% Don't change these lines
\bsp	% typesetting comment
\label{lastpage}
\end{document}